\documentclass[conference]{IEEEtran}
\IEEEoverridecommandlockouts
\usepackage{cite}
\usepackage{amsmath,amssymb,amsfonts}
\usepackage{algorithmic}
\usepackage{graphicx}
\usepackage{textcomp}
\usepackage{xcolor}
\def\BibTeX{{\rm B\kern-.05em{\sc i\kern-.025em b}\kern-.08em
    T\kern-.1667em\lower.7ex\hbox{E}\kern-.125emX}}

\usepackage[ruled,vlined,linesnumbered]{algorithm2e}
\usepackage{multirow}
\usepackage[export]{adjustbox}
\usepackage{xspace}
\usepackage{comment}

\usepackage{amssymb}% http://ctan.org/pkg/amssymb
\usepackage{pifont}% http://ctan.org/pkg/pifont
\definecolor{munsell}{rgb}{0.0, 0.5, 0.69}
\newcommand{\cmark}{\color{munsell}\ding{51}\color{black}}%
\newcommand{\xmark}{\color{red}\ding{55}\color{black}}%
\newcommand{\amark}{\color{black}\ding{86}\color{black}}%

\usepackage[hidelinks]{hyperref}

\usepackage[linesnumbered, ruled]{algorithm2e}
\usepackage{fontawesome5}
\usepackage{subcaption}
\usepackage{makecell}
\usepackage[linesnumbered, ruled]{algorithm2e}

\makeatletter
\let\original@algocf@latexcaption\algocf@latexcaption
\long\def\algocf@latexcaption#1[#2]{%
  \@ifundefined{NR@gettitle}{%
    \def\@currentlabelname{#2}%
  }{%
    \NR@gettitle{#2}%
  }%
  \original@algocf@latexcaption{#1}[{#2}]%
}
\makeatother

\makeatletter
\renewcommand{\@algocf@capt@plain}{above}

\begin{document}

\title{FlexDoc: Flexible Document Adaptation through Optimizing both Content and Layout}
%\xspace
\newcommand{\name}[1]{\def\papername{\textsc{#1}\xspace}}
\name{FlexDoc}
\newcommand{\nameEditor}[1]{\def\papernameEditor{\textsc{#1}\xspace}}
\nameEditor{FlexDocEditor}

%{\footnotesize \textsuperscript{*}Note: Sub-titles are not captured in %Xplore and

\author{
 Yue Jiang\IEEEauthorrefmark{1}, Christof Lutteroth\IEEEauthorrefmark{2}, Rajiv Jain\IEEEauthorrefmark{3}, Christopher Tensmeyer\IEEEauthorrefmark{3}, \\
 Varun Manjunatha\IEEEauthorrefmark{3}, Wolfgang Stuerzlinger\IEEEauthorrefmark{4}, Vlad I. Morariu\IEEEauthorrefmark{3} \\

\IEEEauthorblockA{\IEEEauthorrefmark{1}\textit{Aalto University, Espoo, Finland, yue.jiang@aalto.fi}}
 
\IEEEauthorblockA{\IEEEauthorrefmark{2}\textit{University of Bath, Bath, United Kingdom, cl2073@bath.ac.uk}}

\IEEEauthorblockA{\IEEEauthorrefmark{3}\textit{Adobe Research, College Park, United States, \{rajijain, tensmeye, vmanjuna, morariu\}@adobe.com}}
\IEEEauthorblockA{\IEEEauthorrefmark{4}\textit{Simon Fraser University, Vancouver, Canada, w.s@sfu.ca}}
}

% \author{
% \IEEEauthorblockN{Yue Jiang}
% \IEEEauthorblockA{
% \textit{Aalto University}\\
% Espoo, Finland \\
% yue.jiang@aalto.fi}
% \and
% \IEEEauthorblockN{Christof Lutteroth}
% \IEEEauthorblockA{\textit{University of Bath}\\
% Bath, United Kingdom \\
% cl2073@bath.ac.uk}
% \and
% \IEEEauthorblockN{Rajiv Jain}
% \IEEEauthorblockA{\textit{Adobe Research}\\
% College Park, United States \\
% rajijain@adobe.com}
% \and
% \IEEEauthorblockN{Christopher Tensmeyer}
% \IEEEauthorblockA{\textit{Adobe Research}\\
% College Park, United States \\
% tensmeye@adobe.com}
% \and
% \IEEEauthorblockN{Varun Manjunatha}
% \IEEEauthorblockA{\textit{Adobe Research}\\
% College Park, United States \\
% vmanjuna@adobe.com}
% \and
% \IEEEauthorblockN{Wolfgang Stuerzlinger}
% \IEEEauthorblockA{\textit{Simon Fraser University}\\
% Vancouver, Canada \\
% w.s@sfu.ca}
% \and
% \IEEEauthorblockN{Vlad Morariu}
% \IEEEauthorblockA{\textit{Adobe Research}\\
% College Park, United States \\
% morariu@adobe.com}
% }

\maketitle
\thispagestyle{plain}
\pagestyle{plain}

\begin{abstract}
Designing adaptive documents that are visually appealing across various devices and for diverse viewers is a challenging task. This is due to the wide variety of devices and different viewer requirements and preferences. Alterations to a document's content, style, or layout often necessitate numerous adjustments, potentially leading to a complete layout redesign.
We introduce \papername, a framework for creating and consuming documents that seamlessly adapt to different devices, author, and viewer preferences and interactions. It eliminates the need to manually create multiple document layouts, as \papername\ enables authors to define desired document properties using templates and employs both discrete and continuous optimization in a novel comprehensive optimization process, which leverages automatic text summarization and image carving techniques to adapt both layout and content during consumption dynamically.
Further, we demonstrate \papername\ in real-world scenarios.
\end{abstract}

% \begin{IEEEkeywords}
% % responsive design;
% % authoring tools;
% % user interface
% % hypermedia creation;
% \end{IEEEkeywords}

\section{Introduction}

\renewcommand{\thefootnote}{\fnsymbol{footnote}}
\footnotetext[1]{This work was done in part while the first author was an intern at Adobe.}
\renewcommand*{\thefootnote}{\arabic{footnote}}
\setcounter{footnote}{0}

% need for adaptation
Document layout is important for effective information consumption in applications ranging from print media to digital forms such as web pages and interactive news readers. As device variety increases, a single document needs to adapt to different screen sizes, orientations, and aspect ratios. This variety also increases the effort for document authors and personalization for individual viewers.

% limitations of existing approaches
Existing commercial document creation tools have limitations in generating adaptive documents (Table~\ref{tbl:comparison_commercial}). These arise from system designs being focused on layout capabilities, an author's capabilities, such as being able to code, and a viewer's options during consumption.
From a system perspective, existing tools necessitate manually defined breakpoints or coding to create multiple versions. For instance, Adobe Acrobat Liquid Mode reflows documents to a single-column format but does not generate multi-column formats, and images on smaller screens shrink instead of reflowing. Similarly, CSS struggles with adaptive multi-column layouts and requires specific configurations, such as a fluid grid, to facilitate image flows. Authors using these tools typically need to code and/or manually set breakpoints for diverse layout sizes. Tools like Webflow eliminate coding but still require authors to set breakpoints and lack support for image flows.

Prior work proposed optimization for generating responsive documents without coding or breakpoints. Table~\ref{tbl:comparison_commercial} lists the most related work. Some work~\cite{donovan2014learning, donovan2015designscape, domshlak2000preference, marcotte2011responsive, nebeling2013responsive, jacobs2003adaptive2} used pre-defined alternatives for optimization requiring authors to craft multiple versions of the same content. Such manual specification makes it hard to deal with a whole collection of documents and lacks content personalization; once the document design is finalized, the content remains fixed.
%To eliminate manual specification of alternative elements, some work explored modifying original elements~\cite{borning2000constraint}. 
%Borning et al.~\cite{borning2000constraint} introduced a constraint-based web page layout method, allowing both authors and viewers to add constraints to specify the layout and content. 
%Yet, all such users needed to modify low-level constraints for optimization, which is a nontrivial task. 
%Laine et al.~\cite{laine2021responsive} cropped images, which often led to information loss.
Laine et al.~\cite{laine2021responsive} enabled some personalization by allowing viewers to select important elements displayed in a larger format, but their approach could not alter the layout structure, only adjusting element sizes.
%Laine et al.~\cite{laine2021responsive} addressed the issue of image cropping by avoiding generating cropped images, as it often led to a loss of information. 

%Also, in current approaches, authors determine a document's content and layout, leaving viewers with very limited control over how they can consume a document. Yet, today the variety of screen sizes typically exceeds what authors consider during document creation.
%
 
We propose \papername, a novel approach for dynamically adapting documents to different screen sizes, and author and viewer preferences. Combining discrete and continuous optimization, \papername creates documents with flexible layouts and adaptive content. 
Our system applies image and natural language processing techniques to automatically generate content variations, such as images of varying sizes and aspect ratios and text summaries of varying lengths.
For authors, \papername offers flexible templates editable interactively without coding or specifying breakpoints. For viewers, \papername enables interactive adaptation of document layout and content for optimal consumption.
We evaluate \papername with the following research questions: \textit{RQ1) Do viewers benefit from interactive adapted content at viewing time according to their preferences?} and
 \textit{RQ2) Do authors benefit from document adaptations with layout and content alternatives?} 
Our findings show that document authors can use \papername to edit layout templates while maintaining readability, and viewers benefit from dynamic documents adapting to their preferences.
We present the following main contributions: 
\begin{enumerate}
\item A novel method for generating and optimizing dynamic content and layout to interactively adapt a document to various devices, author preferences, and viewer preferences. Both authors and viewers can influence the layout and content shown in a document.
\item An optimization approach that combines both discrete and continuous optimization of global properties, such as layout and aesthetics, and local properties, such as information loss, content preferences, and interactive adjustments in the level of detail.
\item A demonstration within application scenarios, showing that \papername supports an immersive and interactive approach for reading documents.
\end{enumerate}

\begin{table}[t]
  \resizebox{\linewidth}{!}{
\begin{tabular}{l l *4c}
\hline
\textbf{} &  \textbf{Functionality} & \textbf{\papername} & \textbf{Acrobat Liquid}  & \textbf{CSS}  & \textbf{Webflow}  \\
\hline
\multirow{3}{*}{System} & Adaptive content & \cmark & \xmark & \xmark & \xmark   \\
 & Image flow & \cmark & \xmark & \amark & \xmark  \\
 & Adaptive multi-column & \cmark & \xmark & \xmark & \cmark  \\
\hline
 \multirow{2}{*}{Author}  & No breakpoints required & \cmark & \cmark & \xmark & \xmark  \\
 & No Need to Code & \cmark & \cmark & \xmark  & \cmark  \\
\hline
  \multirow{1}{*}{Viewer}  & Viewer preferences & \cmark & \xmark & \xmark & \xmark  \\
\hline
\end{tabular}}
\caption{Comparative analysis of \papername and existing commercial document tools regarding document adaptation capabilities from system, author, and viewer perspectives. `\cmark' and `\xmark' indicate the presence and absence of functionality. `\amark' denotes functionality achievable through coding.}
\label{tbl:comparison_commercial}
\end{table}

\begin{table}[t]
  \resizebox{\linewidth}{!}{
\begin{tabular}{l p{3cm} *4c}
\hline
\textbf{} &  \textbf{Functionality} & \textbf{\papername} & \textbf{O'Donovan et al.}  & \textbf{Borning et al.}  & \textbf{Laine et al.}  \\
\hline
 \multirow{5}{*}{Author} 
 & No need to create all alternatives manually & \multirow{2}{*}{\cmark} & \multirow{2}{*}{\xmark}& \multirow{2}{*}{\xmark} & \multirow{2}{*}{\cmark}  \\
  & No need to modify low-level constraints & \multirow{2}{*}{\cmark} & \multirow{2}{*}{\cmark} &  \multirow{2}{*}{\xmark}  & \multirow{2}{*}{\cmark} \\
    & No image distortion & \cmark & \cmark & \cmark & \xmark  \\
\hline
  \multirow{4}{*}{Viewer}   & Viewer preferences & \cmark & \xmark & \cmark & \cmark  \\
  & No need to modify low-level constraints & \multirow{2}{*}{\cmark} & \multirow{2}{*}{\cmark}  & \multirow{2}{*}{\xmark}  & \multirow{2}{*}{\cmark}  \\
    & Layout modification & \cmark & \xmark  & \xmark  & \xmark   \\
\hline
\end{tabular} }
\caption{Comparative analysis of \papername and existing document tools from prior research regarding their ability to adapt documents to various authors' and viewers' preferences. '\cmark' and '\xmark ' indicate the presence and absence of functionality.}
\label{tbl:comparison_research}
\end{table}

\section{Document Optimization}
\label{sec:optimization}

Designing an adaptive document involves optimizing content and layout to fit screen properties, author preferences, and viewer preferences. 
%Our objective is to provide a comprehensive method that considers all these aspects. 
This requires numerous element-specific and layout-related decisions. To realize such functionality, we formulate the document optimization problem as a joint discrete and continuous optimization process. Implementation details are available in the supplementary materials.

\subsection{Problem Formulation}

We define the document problem as an optimization problem, where we decide on the positions and sizes of document elements, denoted as ${e}_i = ({x}_i, {y}_i, {w}_i, {h}_i)$. Here, $x_i$ and $y_i$ represent the coordinates of the top-left corner of the $i$-th element, and $w_i$ and $h_i$ represent its width and height, respectively. We focus on rectangular or rectangular bounding box elements without considering hierarchies.

We now define a continuous loss term $\mathcal{L}_\textrm{cont}$ and the discrete loss term $\mathcal{L}_\textrm{disc}$. The continuous loss term focuses on screen and element properties, such as element sizes and overall aesthetics; the discrete loss term focuses on author and viewer preferences. 
The overall objective function is as follows:

\begin{equation}
\begin{split}
\label{eq:objective_function}
&\mathcal{L}(\hat{e}_1, \hat{e}_2, ..., \hat{e}_N, {e}_{p1}, {e}_{p2}, ..., {e}_{pN}; \mathbf{W}_{\mathrm{cont}}, \mathbf{W}_{\mathrm{disc}}) \\
&= \mathcal{L}_{\mathrm{cont}}({\hat{e}_1, \hat{e}_2, ..., \hat{e}_N}, {e}_{p1}, {e}_{p2}, ..., {e}_{pN}; \mathbf{W}_{\mathrm{cont}}) \\
&+ \mathcal{L}_{\mathrm{disc}}(\hat{e}_1, \hat{e}_2, ..., \hat{e}_N; \mathbf{W}_{\mathrm{disc}}),
\end{split}
\end{equation}
where $N$ denotes the total number of elements, $\hat{e}_i = (\hat{x}_i, \hat{y}_i, \hat{w}_i, \hat{h}_i)$ represent the predicted position and size of each GUI element, and $e_{pi} = (x_{pi}, y_{pi}, w_{pi}, h_{pi})$ denote the preferred element positions and sizes. $\mathbf{W}_{\mathrm{cont}}$ is a set of weights assigned to specific continuous properties, and $\mathbf{W}_{\mathrm{disc}}$ consists of weights associated with discrete properties.
We then minimize the objective function to optimize the positions and sizes of document elements:

\begin{equation}
\begin{split}
\{\hat{e}_1, \hat{e}_2, ..., \hat{e}_N\}^{*}
 = \textrm{argmin}_{\{\hat{e}_1, \hat{e}_2, ..., \hat{e}_N\}}& \\ 
\mathcal{L}(\hat{e}_1, \hat{e}_2, ..., \hat{e}_N; &\mathbf{W}_{\mathrm{cont}}, \mathbf{W}_{\mathrm{disc}}).
\end{split}
\label{}
\end{equation}

\begin{figure*}
  \includegraphics[width=0.9\textwidth]{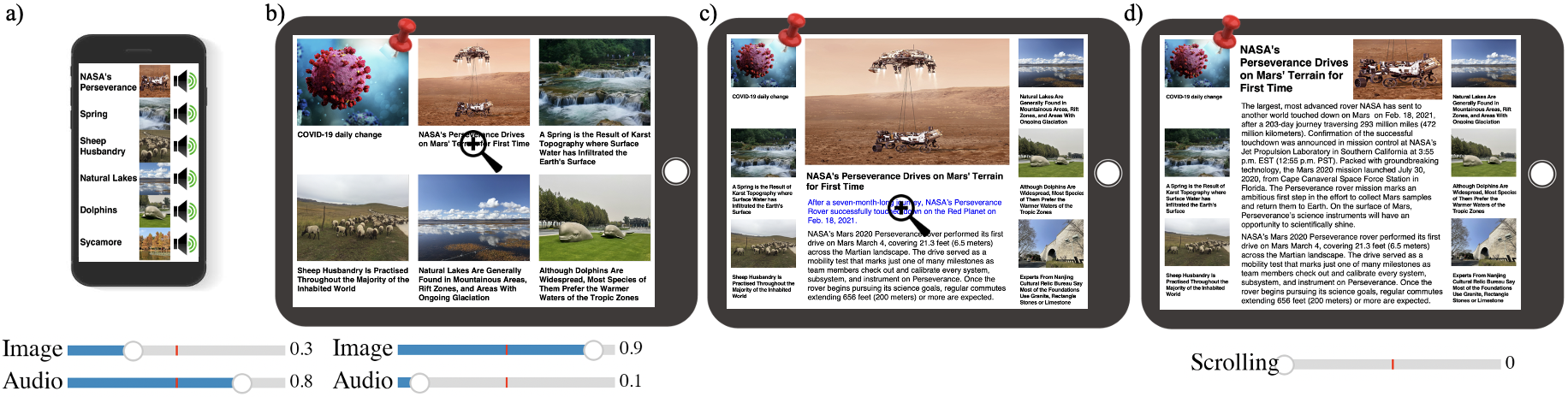}
  \caption[]{
  a) \papername adapts a news page on a mobile phone to provide a compact overview with quick access to audio content for each article, based on viewer preferences (sliders below images), which prioritize audio content over images and text here.
  b) The same news page adapted for a tablet device with a user preference for image content.
  c) As the viewer `pins' the COVID article and `zooms in' on the Mars article, \papername rearranges the layout accordingly, keeping the pinned article in place.
 d) As the viewer `zooms in' on the blue text paragraph in the previous image with a preference for avoiding scrolling, \papername extends the paragraph to provide more details and crops the top image, avoiding the need for scrolling\protect\footnotemark.}
%  \Description{}
  \label{fig:teaser}
\end{figure*}

\subsection{Continuous Loss Term}

The continuous loss term $\mathcal{L}_{\mathrm{cont}}$ specifies the relationship between elements and their properties. Here, we include image loss $\mathcal{L}_{\mathrm{img}}$, text loss $\mathcal{L}_{\mathrm{text}}$, and alignment loss $\mathcal{L}_{\mathrm{align}}$:

\begin{equation}
\begin{split}
\mathcal{L}_{\mathrm{cont}}({\hat{e}_1, \hat{e}_2, ..., \hat{e}_N}, {e}_{p1}, {e}_{p2}, ..., {e}_{pN}; \mathbf{W}_{\mathrm{cont}}) \\
= \mathbf{w}_{\mathrm{img}}\mathcal{L}_{\mathrm{img}} + \mathbf{w}_{\mathrm{text}}\mathcal{L}_{\mathrm{text}} + \mathbf{w}_{\mathrm{align}}\mathcal{L}_{\mathrm{align}},
\end{split}
\end{equation}
where $\mathbf{W}_{\mathrm{cont}} = \{\mathbf{w}_\mathrm{img}, \mathbf{w}_\mathrm{text}, \mathbf{w}_\mathrm{align}\}$ are the weights for images, texts, and alignments, currently set to 1.

\subsubsection{Image Loss}

To optimize an image, we penalize deviations from its preferred size (size loss) and aspect ratio (aspect ratio loss). Directly using the difference in image {\em area} is not advisable since it can significantly distort the image. 

\begin{comment}
    
Thus, we measure image size only based on the difference in size in each dimension. Hence, we define image size loss $\mathcal{L}_{\textrm{s}}$ as 
\begin{equation}
\mathcal{L}_{\textrm{s}} = \dfrac{1}{N_\textrm{img}}\sum_{i=1}^{N_\textrm{img}}|\hat{w}_i - w_{{pi}}|^2 + \dfrac{1}{N_\textrm{img}}\sum_{i=1}^{N_\textrm{img}}|\hat{h}_i - h_{{pi}}|^2,
\end{equation}
%
where ${N_\textrm{img}}$ is the number of images.

Aspect ratio is another important measure for images, defined as $w/h$. The ideal situation is to maintain the same aspect ratio as for the preferred size, {\em i.e.,} $\hat{w}_i /\hat{h}_i  =  w_{{pi}}  /  h_{{pi}} $, which is equivalent to $\hat{w}_i \cdot h_{{pi}} = \hat{h}_i  \cdot w_{{pi}}$. Hence, we define the aspect ratio loss $\mathcal{L}_{\textrm{ar}}$ as 
\begin{equation}
\mathcal{L}_{\textrm{ar}} = \dfrac{1}{{N_\textrm{img}}}\sum_{i=1}^{N_\textrm{img}}|\hat{w}_i \cdot h_{{pi}} - \hat{h}_i  \cdot w_{{pi}}|^2.
\end{equation}
The final image loss is the sum of the size and aspect ratio loss terms. 
$
\mathcal{L}_{\textrm{img}} = \mathcal{L}_{\textrm{s}} + \mathcal{L}_{\textrm{ar}}
$.
\end{comment}

\subsubsection{Text Loss}

We penalize text that is too small to read by considering its size deficit, i.e., by how much its font size $\hat{f}_i$ is smaller than the viewer's preferred font size $f_{pi}$. If the text size exceeds the preferred font size, the size deficit is 0. Our system can dynamically generate shortened versions of text to better fit the document. In such cases, we further penalize text changes if the shortened version is used. 

\begin{comment}

We use the summarization evaluation metric BERTScore~\cite{zhang2019bertscore} to measure the similarity between the shortened version $\hat{t}_i$ and the original version $t_{\textrm{orig}}$ to improve the overall readability. A higher BERTScore indicates greater similarity. Thus, the text loss in \papername is defined as

\begin{equation}
\mathcal{L}_{\textrm{text}} = \dfrac{1}{{N_\textrm{text}}}\sum_{i=1}^{N_\textrm{text}} \max(f_{{pi}} - \hat{f}_i, 0) - \dfrac{1}{{N_\textrm{text}}}\sum_{i=1}^{N_\textrm{text}} \textrm{BERTScore}(\hat{t}_i, t_{\textrm{orig}}),
\end{equation}
%
where ${N_\textrm{text}}$ is the number of text items.
\end{comment}

\subsubsection{Alignment Loss}

We use a measure of the overall aesthetic of a layout based on established visual principles~\cite{scholgens2016aesthetic}. This overall aesthetics loss term could be easily extended to consider additional aesthetic principles. 

\begin{comment}
\papername currently uses alignment loss. For example, we penalize when the images $e_i$ and $e_j$ in the same row are not aligned along the horizontal midline:

\begin{equation}
\mathcal{L}_{\textrm{align}_{(i, j)}} = |(\hat{y}_i + \dfrac{1}{2} \hat{h}_i) -  (\hat{y}_j + \dfrac{1}{2} \hat{h}_j)|^2
\end{equation}

\end{comment}

\subsection{Discrete Loss Term}

The discrete loss term $\mathcal{L}_{\mathrm{disc}}$ involves the selection of templates and individual content alternatives. 
For each element $e_i$, if the viewer has no specific preference, the discrete loss for this element is determined by the author preference loss, $\mathcal{L}_{\mathrm{author}, i}$. When the viewer indicates their preferences without interacting directly, the discrete loss shifts to the viewer preference loss, $\mathcal{L}_{\mathrm{viewer}, i}$. However, if the viewer actively interacts with the content, the discrete loss is governed by the viewer interaction loss, $\mathcal{L}_{\mathrm{int}, i}$, ensuring that the content dynamically adjusts to their direct input.
% It includes author preference loss $\mathcal{L}_{\mathrm{author}}$, viewer preference loss $\mathcal{L}_{\mathrm{viewer}}$, and viewer interaction $\mathcal{L}_{\mathrm{int}}$:

 \begin{equation}
 \begin{split}
 &\mathcal{L}_{\mathrm{disc}}({\hat{e}_1, \hat{e}_2, ..., \hat{e}_N}; \mathbf{W}_{\mathrm{disc}}) \\
 &= \sum_i \mathbf{w}_{\mathrm{author}, i}\mathcal{L}_{\mathrm{author}, i} + \mathbf{w}_{\mathrm{viewer}, i}\mathcal{L}_{\mathrm{viewer}, i} + \mathbf{w}_{\mathrm{int}, i}\mathcal{L}_{\mathrm{int}, i},
\end{split}
 \end{equation}
% %
where one of $\{\mathbf{w}_{\mathrm{author}, i}$, $\mathbf{w}_{\mathrm{viewer}, i}$, $\mathbf{w}_{\mathrm{int}, i}\}$ is 1 and the other two are 0, depending on whether the viewer sets their preferences or interacts with the content.

\subsubsection{Author Preference Loss}

Document authors can define alternatives for both layout templates and content, each assigned preference ranks. Higher loss values are assigned to lower-ranked templates. Specifically, the $m$-th ranked template is assigned a loss value of $-1000\cdot (M + 1 - m)$, prioritizing more preferred templates, where $M$ is the number of template alternatives. This approach creates a gradient of loss values across ranks, allowing for optimization within a template before transitioning to another. 

\begin{comment}
Similarly, for the $i$-th content, the $k$-th ranked template is assigned a loss value of $-50\cdot (K_i + 1 - k_i)$ to prioritize more preferred alternatives, where $K$ is the number of alternatives for that specific content. The final loss for the author's preference is then calculated as follows:

\begin{equation}
\mathcal{L}_{\textrm{author}} = -1000\cdot (M + 1 - m) - \sum_{i=1}^N 50\cdot (K_i + 1 - k_i).
\end{equation}
\end{comment}

\subsubsection{Viewer Preferences}
Viewer preferences have higher priorities than those specified by authors, as the end goal of our approach is to enhance the viewing experience. As shown in \autoref{fig:teaser}, viewers can adjust their preferences with the sliders. For example, if the viewer increases their preference for ``images'' using the corresponding slider, the loss value of all other alternatives will be decreased so that images are more likely to be chosen. 

\begin{comment}
The range of sliders is within [0, 1], where 0.5 indicates no change in preferences. We denote the slider value to be $s_k$. Other loss values remain the same as defined by author preferences if not set by the viewer. Thus, the viewer preference loss is defined as 

\begin{equation}
\mathcal{L}_{\textrm{viewer}} = \sum_{i=1}^N (0.5 - s_k)\cdot 50\cdot (K_i + 1 - k_i)).
\end{equation}
\end{comment}

\subsubsection{Viewer Interactions}
\papername can dynamically change the screen's content in response to viewer interactions. Viewer interactions are given the highest priority since they represent direct requests from the user. Thus, if the viewer chooses a specific template or content alternative through an interaction, that alternative must be selected unless no solution exists. Other contents are then optimized accordingly.

\subsection{Dynamic Content Generation}
\label{sec:dynamiccontent}

 \begin{comment}
\begin{figure}[t]
\centering
\includegraphics[width=\linewidth]{figures/content_generation.png}
\caption{Examples of image seam carving and content summarization.}
%\Description{Examples of image seam carving and content summarization.}
\label{fig:generation}
\end{figure}

\end{comment}
 
Given a screen/window size, we optimize the positions and sizes of elements and alternative selection. However, fitting content into a layout is often challenging, especially for smaller screen sizes. 
%One solution is to scale images and text. However, this would reduce content readability if images or fonts become too small. Alternatively, we could maximize readability by cutting content, but then we would lose (too) much information. 
To accommodate the diversity of screen sizes and document author and viewer preferences, \papername dynamically selects or generates alternative content. It applies seam carving for image adaptation and BERT-based text summarization for variable-sized texts. It then optimizes across the potential alternatives with the given screen size and viewer preferences.
Further details are in the supplementary materials.

\footnotetext{Image credits: Production Perig/stock.adobe.com and NASA/JPL-Caltech}

\section{Document Authoring and Viewing}
\label{sec:authoringviewing}

% \begin{figure*}[t]
% \centering
% \includegraphics[width=\textwidth]{figures/framework.pdf}
% \caption{
% The layout template can be extracted through reverse engineering of an existing document or created directly by the author. Subsequently, the document can then be optimized based on the screen size and author preferences. Once the screen size changes and/or the viewer indicates their preferences, the document can adapt accordingly.
% }
% %\Description{}
% \label{fig:framework}
% \end{figure*}

\papername optimizes the document based on the screen properties, author preferences, and viewer preferences. 
%As shown in \autoref{fig:framework}, the layout template can be extracted through reverse engineering of an existing document or created directly by the author.
Authors can define their content preferences using \papernameEditor, a graphical document template editor. This editor allows \emph{authors} to guide the optimization process by providing different layout templates, content alternatives, and preference rankings. 
Subsequently, the document can then be optimized based on the screen size and author preferences. 
On the other hand, \emph{viewers} can also adapt a document by selecting different layout templates or adjusting their preferences via simple operations. %sliders (e.g., for different content modalities).%, or interacting directly with the content. 
Once the screen size changes and/or the viewer changes their preferences, the document can adapt accordingly.
%These changes are saved and applied to other documents, reducing the need for repeated adjustments and ensuring a consistent viewing experience. 
More details are in the supplementary materials.

\section{Applications}
\label{sec:application}

We demonstrate \papername in multiple real-world application scenarios. Here, we show an example of news reading. Other examples are in the supplementary materials.

\begin{figure*}[t!]
\centering
\includegraphics[width=0.7\textwidth]{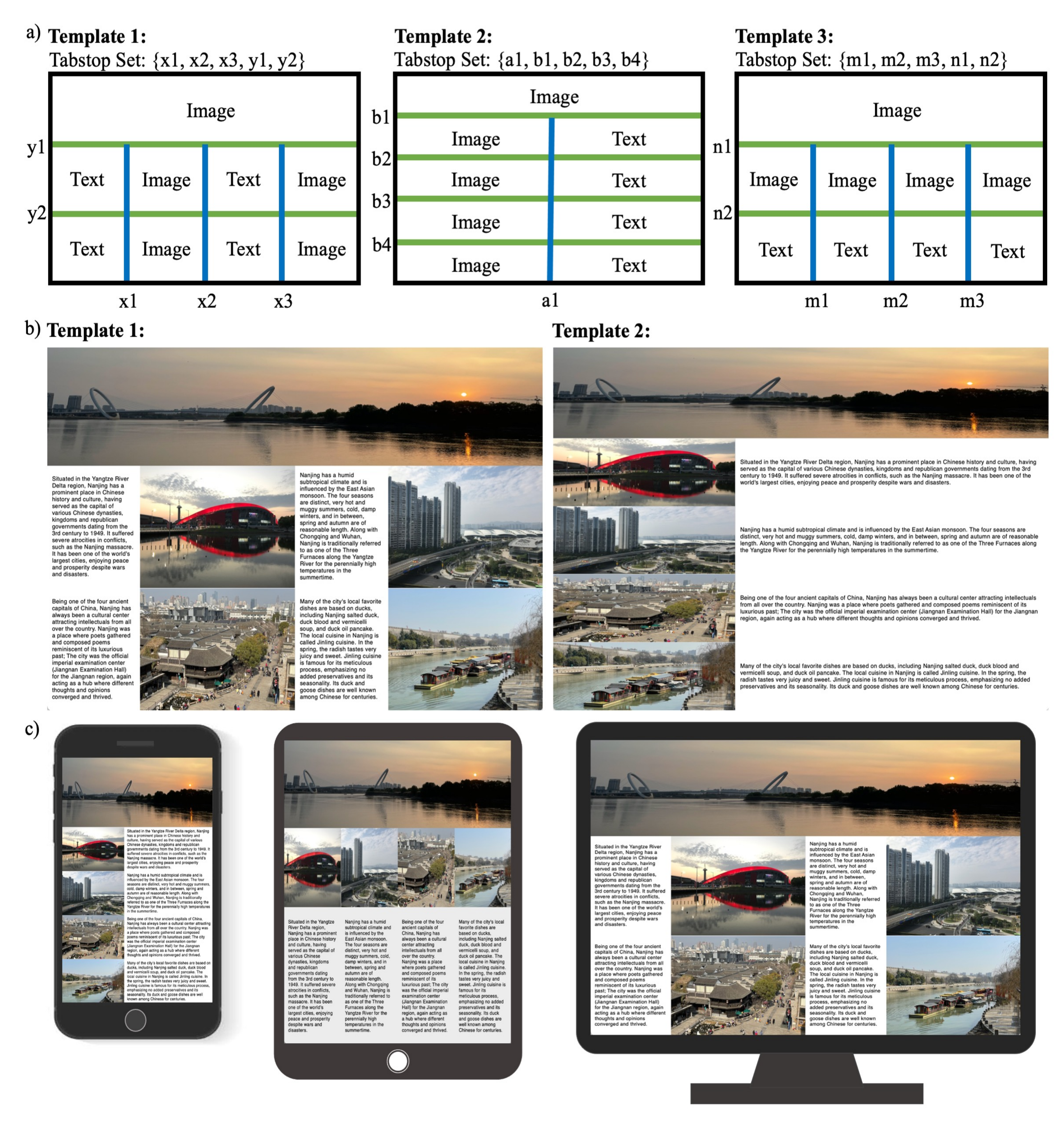}
\caption{Document optimization results: a) The author defines three templates. \papername optimizes tabstop positions based on the objective function. b) Document results when the viewer prefers the first or second template, respectively. c) Results on different devices, which balance both layout structure and the amount of content.
}
%\Description{This figure illustrates the document optimization process: a) Each template is defined as a set of tabstops, which are grid lines defining the boundaries of the document elements; b) shows that \papername generates new images and text by image seam carving and content summarizing; c) demonstrates that the result balances both layout structure and the amount of content.}
\label{fig:result}
\end{figure*}

Viewers have different preferences for news consumption, based on personal interests and desired levels of detail. Modern news websites use location and browsing history to recommend and preview news items on the front page. These previews typically highlight critical information within a concise format, without comprehensive details and background context.
Further, individual news items often mention only the latest events without reference to previous news messages in the series or background information. This requires viewers who want more detail or background information to search for related documents or follow links provided in the document they are reading. Instead of redirecting to other documents, viewers could be better served by extending the document (using content from related articles, not AI-generated ones) they are reading based on their needs, generating more detailed information within the document. %Note that extending the document does not refer to AI-generated content but to adding the content from related articles.

\subsubsection{Front Page Optimization}

%News front pages are continuously updated as the latest news comes in. However, viewers cannot control how the front page displays all news items. For example, updated news often replaces previous items, making it harder for the viewer to find such previous news items. It would be helpful to be able to pin news of continual interest to a given viewer, such as stock market or COVID-19 news, which could then continuously show follow-up related news.

News front pages constantly update with the latest news, which can replace older items and give viewers access to previous articles. Allowing viewers to ``pin'' their interests, like stock market or COVID-19 news, with \papername enables continuous access to related updates.

\papername enables viewers to adapt a news front page to their individual preferences (Figure~\ref{fig:teaser}a and b).
%show the front pages for two different viewers, each based on their respective preferences. 
Viewers can choose preferred modalities and detail levels, and `pin' items of interest in place. This approach gives freedom to both authors and viewers, influencing the final content and layout. Authors benefit from \papername automating much of the document adaptation work. Authors predefine layout and content alternatives, allowing viewers to finalize choices. Viewers can adjust reading preferences interactively, like `zooming in' on specific news without losing the overall context of the news front page (Figure~\ref{fig:teaser}c). 

\subsubsection{Dynamic Documents}

\papername aims to generate different aesthetically pleasing document alternatives automatically and dynamically based on a viewer's personal preferences. Unlike news recommendation websites like Google News, which essentially only reorder news items based on viewer interests, \papername enhances the reading experience by dynamically extending and shortening document content on demand. For instance, a viewer can `zoom in' on a part of an article to get more detail, and \papername then automatically re-optimizes the layout to accommodate this extra detail (Figure~\ref{fig:teaser}d).

\section{Evaluation}

To understand the benefits and challenges of \papername, we examine how different users might use \papername from both author and viewer perspectives.
%, we conducted an evaluation centered on two research questions: 
% RQ1) Do authors benefit from document adaptations enabled by providing layout and content alternatives? 
% and RQ2) Do viewers benefit from content that they can adapt interactively at viewing time according to their preferences?

\subsubsection{Participants}
We interviewed 13 interface designers (6M, 7F) including 10 \textit{Professional Designers} with over 2 years of professional UI/UX design experience in industry or research labs; and 3 \textit{Non-Professional Designers} who are HCI graduate students with some interface prototyping experience. Five participants were interviewed in person, and the others were interviewed remotely through video conferencing. 

\subsubsection{Materials}

Participants used laptops for authoring documents with \papernameEditor and for viewing them.  

\subsubsection{Experiment Design}

The study used a within-subject design, requiring participants to compare the use of \papername and the existing document tools they normally use.

\subsubsection{Procedure}

After explaining the basic ideas of \papername, participants could experience \papername from both the author's and the viewer's perspectives through three tasks: a) Participants used the \papernameEditor to create a news website resembling Figure~\ref{fig:teaser}. They then interacted with the created news website and compared it with their experience using other news websites like Google News. b) Participants used \papernameEditor to create their own document templates and assigned content, observing how \papername adapts these documents to different devices and viewer-preferred templates. They then compared \papername with their usual design tools for document creation. c) Focusing on the viewer's perspective, participants experienced how \papername adapts a draft of the \papername paper to different devices and formats. They compared this with their previous experience of reading papers or similar documents on different devices. 
% We recorded and transcribed participants' feedback. % on both the authoring (RQ1) and viewing experience (RQ2).

\subsubsection{Findings}
We perform qualitative analysis from both the author's and viewer's perspectives.

\textit{RQ1: Do viewers benefit from interactive adapted content at viewing time according to their preferences?}

% dynamic level of detail
\textit{Viewers benefit from the dynamic content generation.}
%All participants commented favorably on the \papername reading experience. 
Participants were most excited by the dynamic level of detail provided, which facilitated easier access to desired content. (\textit{``It happened a lot that right after I opened the news article, I realized that I am not interested []. I really like the idea of showing [something like an] abstract before opening the article.''} (P1), \textit{``I can get the suitable amount of information I need. It will save me a lot of time.''} (P6), \textit{``see the summary and being able to hop around''} (P13)).
% flexibility

\textit{Flexibility to adapt to different devices and viewer preferences.}
The second-most mentioned advantage of \papername was its ability to adjust the document layout to different devices (\textit{``Everything becomes almost unreadable on my phone. This image reflow function solves the problem.''} (P6), \textit{``Google News is not optimized very well for mobile devices''} (P3)).
% comfort & accessibility
%Finally, most liked the idea of adapting content to make the reading experience more comfortable and accessible (\textit{``it is a big step forward that readers could define their comfortable layout for reading''} (P3), \textit{``Accessibility for different types of users that require bigger text or more contrast''} (P11), \textit{``make it different from if I was teaching sixth graders to fourth graders''} (P11)).
In summary, all participants identified tangible benefits due to their flexible reading experience with \papername.

\textit{RQ2: Do authors benefit from document adaptations with layout and content alternatives?}

% flexibility required
\textit{Reduced design effort to create responsive documents.}
Most participants emphasized that designing content for a variety of formats and sizes is a common requirement today (\textit{``In the future of digital publishing authors cannot account for all the screen sizes they need to account for, and designers are pressured to produce content for more and more different surfaces''} (P11)).
% \papername editor is more powerful
Some participants noted that \papernameEditor offers greater capabilities compared to their current design tools, particularly in providing flexible layouts for different devices (\textit{``XD does this in a simplistic way, which is more about devices... [\papername{}] is much more powerful''} (P11)).
% ease of use & high-level layout editing
Despite the added authoring complexity, participants found \papernameEditor fairly easy to use (\textit{``For people who are not a software developer, the template creation looks intuitive and easy to do.''} (P5)).  They appreciated how it allowed them to work at a higher level of abstraction in designing document layouts(\textit{``Designers often cannot think about the overall layout. Designers mostly focus on the component perspective and often ignore the overall layout. This system fills [] this gap.''} (P4), \textit{``don't need to get to be too specific about something like alignment. I think alignment is generally time-consuming to deal with ... spent so much time on those small issues. So I think this system can help them avoid those issues. It is good for overall responsive, adaptive layout creation.''} (P6)).
% reduced design effort with pre-defined templates 
Furthermore, many participants recognized that templates could save them time: \textit{``I think this kind of templates help me solve the alignment issue. I think it can significantly reduce the design effort.''} (P4), \textit{``I hope to have some predefined templates so that I don't need to think about how to design the template myself.''} (P1), and \textit{``The general idea of using templates is cool and very convenient.''} (P2).
This overall positive reception indicates that participants recognized the benefits that \papername can provide for document authors and that they valued the ease of creating adaptive content.

%\paragraph{Limitations}

% challenge of previewing behaviour
\textit{Lack of functionality to preview resize behaviors.}
Some participants noted that while it was fairly easy to create flexible documents, it was less straightforward to understand their layout resize behavior across the many possible sizes (\textit{``from the designer’s point of view you almost need a simulator to show me what my design is going to look like...you can’t show me everything...the challenges is do the designers have the ability to preview the results''} (P11)).

% challenge of template editing
\textit{Template editing can be challenging for less experienced designers.}
Some participants mentioned that editing \papername templates might require technical expertise that not every designer had (\textit{``would probably be intimidating for an everyday commonplace user, so being able to make it less technical looking might help''} (P12), \textit{``There are many different layout problems that can come up with this, so just the interactions with those decisions might require more testing''} (P11)).

\section{Discussion and Future Work}

%Our proposed \papername approach dynamically optimizes both the structure and content of a document to adapt it to various device properties and the preferences of authors and viewers. \papername can be applied to a wide range of document-centric applications, enhancing both the reading and authoring experience across various document types. Designers can use \papernameEditor to create flexible document layout templates that ensure readability across different use cases. \papername can then generate suitable versions of both images and text content to fit these different layout templates. \papername can adapt UIs in interactive times on a laptop with an Intel i5 CPU (less than half a second to optimize for 100 elements). It can be applied to any PDF document or webpage, through detecting the element types and bounding boxes using document object detection methods~\cite{li2020cross}. 

Our \papername approach dynamically optimizes document structure and content to adapt to various devices and user preferences. Applicable to a wide range of document-centric applications, \papername enhances both reading and authoring experiences. Designers can use \papernameEditor to create flexible layouts that ensure readability across different use cases. \papername generates suitable versions of images and text to fit these layouts and can adapt UIs interactively in under half a second on an Intel i5 laptop. It can be applied to any PDF or webpage by detecting element types and bounding boxes using document object detection methods~\cite{li2020cross}.

\papername does not currently consider semantic relationships or hierarchy among document elements, nor handles elements with irregular boundaries. Future work could extend to elements with irregular boundaries and optimize based on document semantics. Additionally, the lack of standard metrics for evaluating adaptive UIs makes quantitative comparison difficult. Future work can establish such metrics.

\section{Related Work}

\renewcommand{\thefootnote}{\fnsymbol{footnote}}
\footnotetext[1]{This work was done in part while the first author was an intern at Adobe.}
\renewcommand*{\thefootnote}{\arabic{footnote}}
\setcounter{footnote}{0}

%This section focuses on customized layout generation, the limitations of preexisting adaptive document approaches, and constraint-based layouts.

\subsection{Layout Alternatives and Customized Layout Generation}

Previous work has proposed optimization-based approaches for customized layout generation to improve the viewer experience across different screen sizes and viewer requirements~\cite{weld2003automatically, fogarty2003gadget, jiang2022computational, jiang2023future, jiang2024computational, jiang2024computational2, jiang2024graph4gui, hegemann2023computational}. SUPPLE~\cite{Gajos2008decision, gajos2004supple} automatically optimized user interfaces by applying alternative widgets or groupings to accommodate screen-size constraints and customize user interfaces for people with disabilities~\cite{gajos2008improving}. Personalization was further improved by specifying a cost function to meet users' preferences and target devices~\cite{Gajos2004improving} and maintaining consistency~\cite{gajos2005cross}. Arnauld~\cite{Gajos2005preference} generated optimized parameters for layouts based on a cost function, resulting in more optimal document layouts.
%and Kido et al.~\cite{kido2015document} used paraphrases to reduce paragraph size when formatting issues were detected, 
Recent work has explored layout design using program synthesis techniques. Scout~\cite{swearngin2020scout} enabled designers to explore alternatives and receive design feedback by generating potential layouts based on user-provided high-level constraints. Yet, Scout only provided fixed-size layout suggestions without dynamic resize behaviors nor guaranteed diverse results. InferUI~\cite{bielik2018robust} inferred constraints to describe a layout from layout examples, but only with linear constraints expressing relative mutual alignments of widgets and a single topological arrangement. It could not handle dynamic topological layouts such as flows and alternative positions. In contrast, \papername can handle layouts containing textflows with proper resize behaviors and dynamic topology and can adapt the document to screen size and viewer preferences.

%For mobile apps and responsive web design, authors typically define several layout alternatives and dynamically select among them to support a variety of devices with different screen sizes, resolutions, and aspect ratios~\cite{sahami2013insights, zeidler2017automatic, marcotte2011responsive}. However, manually specifying multiple layout alternatives can take much time and effort, resulting in duplicate work and potential dependencies/discrepancies between alternatives. Some automatic layout generation approaches used templates or modifiable suggestions~\cite{schrier2008adaptive, sinha2015responsive, zanden1990automatic}, allowing document authors to choose from and modify the generated recommendations. Similarly, \papername uses adaptive layout templates to create layout alternatives but can do this at runtime.

\subsection{Adaptive Documents}

Early document layout builders focused on document architecture and formatting to arrange text into lines, paragraphs, and other high-level structures~\cite{peels1985document, knuth1981breaking, furuta1982document}. LaTeX uses a dynamic programming approach to solve the problem of breaking a paragraph into lines~\cite{mittelbach2000formatting}. Later work generated adaptive web and document layouts by varying document representations. Chen et al.~\cite{chen2005adapting} proposed adapting web pages for small screen devices by dividing the original page into smaller blocks. Xie et al.~\cite{xie2005adaptive} presented a novel tree representation for displaying documents on various screens. FrameKit \cite{wu2024framekit} generates resizable UIs via keyframe interpolation. Domshlak et al.~\cite{domshlak2000preference} proposed a system for personalized presentation and a preference-based configuration process for web pages. Their approaches also optimized the selection of document content alternatives based on the author's and the viewer's preferences. However, all the alternatives were discrete and predefined. In contrast, in addition to predefined content alternatives, \papername can automatically generate new ones and further optimize the selected alternatives to fit better into the space available for the document in the current context.

Recent document layout generation research used deep learning approaches to avoid manually defining constraints and templates. LayoutGAN~\cite{li2019layoutgan} proposed a generative model to place graphics elements into a document layout. LayoutGAN++~\cite{kikuchi2021constrained} improved the generative layout model with transformer blocks and latent space exploration. Zheng et al.~\cite{zheng2019content} added a content-awareness factor to generate document layouts based on the document topics. Neural Design Network~\cite{lee2019neural} generated document layouts via deep learning networks to produce layouts that follow given constraints. However, deep learning approaches can only produce the document styles represented in their training data, which typically biases their outputs. Furthermore, they give authors less control over the generated documents and cannot generate documents that adapt dynamically, {\em i.e.,} to the current window/screen size. In contrast, \papername gives control over the documents to both authors and viewers and, at the same time, automatically generates proper text and images to fit the screen better.

 \subsection{Constraint-based Resizable Layout}

The need for resizable layouts is driven by the vast diversity of screen sizes and aspect ratios of current devices and the ability of desktop graphical user interfaces to resize windows interactively. While early layout models such as group, grid, table, and grid-bag layouts~\cite{myers2000past, myers1995user, myers97theamulet} provided basic functionality, more recent constraint-based layout models~\cite{zeidler2017tiling, lutteroth2008domain} offer advanced options for generating resizable and responsive layouts~\cite{karsenty1993inferring, scoditti2009new, weber2010reduction, zeidler2012auckland}.
Recent work on a more expressive layout model for graphical user interfaces (GUIs), called ORC Layout~\cite{jiang2019orclayout}, unifies flow layouts and conventional constraint-based layouts through OR-constraints (ORC). OR-constraints are a combination of hard and soft constraints, where the entire OR-constraint is a hard constraint, and each disjunctive part is a soft constraint. This allows for the definition of alternatives for layout components, enabling a single layout specification to create adaptive layouts for a wide range of screen sizes, orientations, and aspect ratios. ORC layout specifications for GUIs can be efficiently solved using ORCSolver~\cite{jiang2020orcsolver} and reverse-engineered from interfaces through ReverseORC~\cite{jiang2020reverseorc}. \papername applies OR-constraints to optimize the layout of adaptive documents, combined with methods for generating content alternatives, to jointly optimize layout and content.
 \section{Document Optimization}
\label{sec:optimization}

Designing an adaptive document involves optimizing both content and layout to fit the screen's properties, author preferences, and viewer preferences. Our objective is to provide a comprehensive method that considers all these aspects. This involves making a large number of both element-specific and layout-related decisions. To achieve this, we formulate the document optimization problem as a joint discrete and continuous optimization process.

\subsection{Problem Formulation}

We define the document problem as an optimization problem. With this formulation, we decide on the positions and sizes of document elements (denoted as ${e}_i = ({x}_i, {y}_i, {w}_i, {h}_i)$, where the coordinates $(x_i, y_i)$ represent the top-left corner of the $i$-th element and $(w_i, h_i)$ represents its width and height), along with the layout of the document. Here, we focus on a setting where all the elements are rectangular or in rectangular bounding boxes, and we do not consider hierarchies.

We define two objective terms: the continuous loss term $\mathcal{L}_\textrm{cont}$ and the discrete loss term $\mathcal{L}_\textrm{disc}$. The continuous loss term focuses on the screen and element properties, such as element sizes and overall aesthetics; the discrete loss term focuses more on the author preferences and viewer preferences. 
The overall objective function is defined as follows:

\begin{equation}
\begin{split}
\label{eq:objective_function}
&\mathcal{L}(\hat{e}_1, \hat{e}_2, ..., \hat{e}_N, {e}_{p1}, {e}_{p2}, ..., {e}_{pN}; \mathbf{W}_{\mathrm{cont}}, \mathbf{W}_{\mathrm{disc}}) \\
&= \mathcal{L}_{\mathrm{cont}}({\hat{e}_1, \hat{e}_2, ..., \hat{e}_N}, {e}_{p1}, {e}_{p2}, ..., {e}_{pN}; \mathbf{W}_{\mathrm{cont}}) \\
&+ \mathcal{L}_{\mathrm{disc}}(\hat{e}_1, \hat{e}_2, ..., \hat{e}_N; \mathbf{W}_{\mathrm{disc}}),
\end{split}
\end{equation}

where the total number of elements is $N$, the predicted position and size of each GUI element as $\hat{e}_i = (\hat{x}_i, \hat{y}_i, \hat{w}_i, \hat{h}_i)$, and the preferred element positions and sizes are represented by $e_{pi} = (x_{pi}, y_{pi}, w_{pi}, h_{pi})$. $\mathbf{W}_{\mathrm{cont}}$ is a set of weights assigned to specific continuous properties, and $\mathbf{W}_{\mathrm{disc}}$ consists of weights associated with individual discrete properties.

We then minimize the objective function to optimize the positions and sizes of document elements.
The optimization process can be represented as

\begin{equation}
\begin{split}
\{\hat{e}_1, \hat{e}_2, ..., \hat{e}_N\}^{*}
 = \textrm{argmin}_{\{\hat{e}_1, \hat{e}_2, ..., \hat{e}_N\}}& \\ 
\mathcal{L}(\hat{e}_1, \hat{e}_2, ..., \hat{e}_N; &\mathbf{W}_{\mathrm{cont}}, \mathbf{W}_{\mathrm{disc}}).
\end{split}
\label{}
\end{equation}

\subsection{Continuous Loss Term}

The continuous loss term $\mathcal{L}_{\mathrm{cont}}$ addresses the relationship between elements and their properties. Here, we include image loss $\mathcal{L}_{\mathrm{img}}$, text loss $\mathcal{L}_{\mathrm{text}}$, and alignment loss $\mathcal{L}_{\mathrm{align}}$:

\begin{equation}
\begin{split}
\mathcal{L}_{\mathrm{cont}}({\hat{e}_1, \hat{e}_2, ..., \hat{e}_N}, {e}_{p1}, {e}_{p2}, ..., {e}_{pN}; \mathbf{W}_{\mathrm{cont}}) \\
= \mathbf{w}_{\mathrm{img}}\mathcal{L}_{\mathrm{img}} + \mathbf{w}_{\mathrm{text}}\mathcal{L}_{\mathrm{text}} + \mathbf{w}_{\mathrm{align}}\mathcal{L}_{\mathrm{align}},
\end{split}
\end{equation}

where $\mathbf{W}_{\mathrm{cont}} = \{\mathbf{w}_\mathrm{img}, \mathbf{w}_\mathrm{text}, \mathbf{w}_\mathrm{align}\}$ are the weights for images, texts, and alignments. We currently set all these to 1.

\subsubsection{Image Loss}

To optimize an image, we penalize deviations from an image's preferred size (size loss) and aspect ratio (aspect ratio loss). We cannot use the difference in image {\em area} directly since this can significantly distort the image by changing its aspect ratio. Thus, we measure image size only based on the difference in size in each dimension. Hence, we define image size loss $\mathcal{L}_{\textrm{s}}$ as 
\begin{equation}
\mathcal{L}_{\textrm{s}} = \dfrac{1}{N_\textrm{img}}\sum_{i=1}^{N_\textrm{img}}|\hat{w}_i - w_{{pi}}|^2 + \dfrac{1}{N_\textrm{img}}\sum_{i=1}^{N_\textrm{img}}|\hat{h}_i - h_{{pi}}|^2,
\end{equation}

where ${N_\textrm{img}}$ is the number of images.

Aspect ratio is another important measure for images, defined as $w/h$. The ideal situation is to maintain the same aspect ratio as for the preferred size, {\em i.e.,} $\hat{w}_i /\hat{h}_i  =  w_{{pi}}  /  h_{{pi}} $, which is equivalent to $\hat{w}_i \cdot h_{{pi}} = \hat{h}_i  \cdot w_{{pi}}$. Hence, we define the aspect ratio loss $\mathcal{L}_{\textrm{ar}}$ as 
\begin{equation}
\mathcal{L}_{\textrm{ar}} = \dfrac{1}{{N_\textrm{img}}}\sum_{i=1}^{N_\textrm{img}}|\hat{w}_i \cdot h_{{pi}} - \hat{h}_i  \cdot w_{{pi}}|^2.
\end{equation}
The final image loss is the sum of the size and aspect ratio loss terms. 
$
\mathcal{L}_{\textrm{img}} = \mathcal{L}_{\textrm{s}} + \mathcal{L}_{\textrm{ar}}
$.

\subsubsection{Text Loss}

We penalize text that is too small to read by considering its size deficit, i.e., by how much its font size $\hat{f}_i$ is smaller than the viewer's preferred font size $f_{pi}$. If the text size is larger than the preferred font size, then the size deficit is 0. Our system can dynamically generate shortened-version texts to fit the document better. In this case, we further penalize text changes if the shortened version is used. We use the summarization evaluation metric BERTScore~\cite{zhang2019bertscore} to measure the similarity between the shortened version $\hat{t}_i$ and the original version $t_{\textrm{orig}}$ to improve the overall readability. A higher BERTScore indicates greater similarity. Thus, the text loss in \papername is defined as

\begin{equation}
\mathcal{L}_{\textrm{text}} = \dfrac{1}{{N_\textrm{text}}}\sum_{i=1}^{N_\textrm{text}} \max(f_{{pi}} - \hat{f}_i, 0) - \dfrac{1}{{N_\textrm{text}}}\sum_{i=1}^{N_\textrm{text}} \textrm{BERTScore}(\hat{t}_i, t_{\textrm{orig}}),
\end{equation}

where ${N_\textrm{text}}$ is the number of text items.

\subsubsection{Alignment Loss}

Previous work explored how to measure the overall aesthetics of a layout based on established visual principles~\cite{scholgens2016aesthetic}. This overall aesthetics loss term could be easily extended to consider different aesthetic principles. \papername currently uses alignment loss. For example, we penalize when the images $e_i$ and $e_j$ in the same row are not aligned along the horizontal midline:

\begin{equation}
\mathcal{L}_{\textrm{align}_{(i, j)}} = |(\hat{y}_i + \dfrac{1}{2} \hat{h}_i) -  (\hat{y}_j + \dfrac{1}{2} \hat{h}_j)|^2
\end{equation}

\subsection{Discrete Loss Term}

The discrete loss term $\mathcal{L}_{\mathrm{disc}}$ involves the selection of templates and individual content alternatives. 
For each element $e_i$, if the viewer has no specific preference, the discrete loss for this element is determined by the author preference loss, $\mathcal{L}_{\mathrm{author}, i}$. When the viewer indicates their preferences without interacting directly, the discrete loss shifts to the viewer preference loss, $\mathcal{L}_{\mathrm{viewer}, i}$. However, if the viewer actively interacts with the content, the discrete loss is governed by the viewer interaction loss, $\mathcal{L}_{\mathrm{int}, i}$, ensuring that the content dynamically adjusts to their direct input.
% It includes author preference loss $\mathcal{L}_{\mathrm{author}}$, viewer preference loss $\mathcal{L}_{\mathrm{viewer}}$, and viewer interaction $\mathcal{L}_{\mathrm{int}}$:

 \begin{equation}
 \begin{split}
 &\mathcal{L}_{\mathrm{disc}}({\hat{e}_1, \hat{e}_2, ..., \hat{e}_N}; \mathbf{W}_{\mathrm{disc}}) \\
 &= \sum_i \mathbf{w}_{\mathrm{author}, i}\mathcal{L}_{\mathrm{author}, i} + \mathbf{w}_{\mathrm{viewer}, i}\mathcal{L}_{\mathrm{viewer}, i} + \mathbf{w}_{\mathrm{int}, i}\mathcal{L}_{\mathrm{int}, i},
\end{split}
 \end{equation}
% %
where one of $\{\mathbf{w}_{\mathrm{author}, i}$, $\mathbf{w}_{\mathrm{viewer}, i}$, $\mathbf{w}_{\mathrm{int}, i}\}$ is 1 and the other two are 0, depending on whether the viewer sets their preferences or interacts with the content.

\subsubsection{Author Preference Loss}

Document authors have the flexibility to define alternatives for both layout templates and content, each associated with preference ranks. Larger loss values are assigned to lower-ranked templates. In practical terms, the $m$-th ranked template is assigned a loss value of $-1000\cdot (M + 1 - m)$, prioritizing more preferred templates, where $M$ is the number of template alternatives. This approach creates a gradient of loss values between ranks, allowing for optimization within a template before transitioning to another. Similarly, for the $i$-th content, the $k$-th ranked template is assigned a loss value of $-50\cdot (K_i + 1 - k_i)$ to prioritize more preferred alternatives, where $K$ is the number of alternatives for that specific content. The final loss for the author's preference is then calculated as follows:

\begin{equation}
\mathcal{L}_{\textrm{author}} = -1000\cdot (M + 1 - m) - \sum_{i=1}^N 50\cdot (K_i + 1 - k_i).
\end{equation}

\subsubsection{Viewer Preferences}
Viewer preferences have higher priorities than those specified by authors, as the end goal of our approach is to enhance the viewing experience. As shown in \autoref{fig:teaser}, viewers can adjust their preferences with the sliders. For example, if the viewer increases their preference for ``images'' using the corresponding slider, the loss value of all image alternatives will be decreased so that images are more likely to be chosen. The range of sliders is within [0, 1], where 0.5 indicates no change in preferences. We denote the slider value to be $s_k$. Other loss values remain the same as defined by author preferences if not set by the viewer. Thus, the viewer preference loss is defined as 

\begin{equation}
\mathcal{L}_{\textrm{viewer}} = \sum_{i=1}^N (0.5 - s_k)\cdot 50\cdot (K_i + 1 - k_i)).
\end{equation}

\subsubsection{Viewer Interactions}
\papername can dynamically change the screen's content in response to viewer interactions. Viewer interactions are given the highest priority since they represent direct requests from the user. Thus, if the viewer chooses a specific template or content alternative through an interaction, that alternative must be selected unless no solution exists. Other contents are then optimized accordingly.

\section{Dynamic Content Generation}
\label{sec:dynamiccontent}
 
\begin{figure}[t]
\centering
\includegraphics[width=\linewidth]{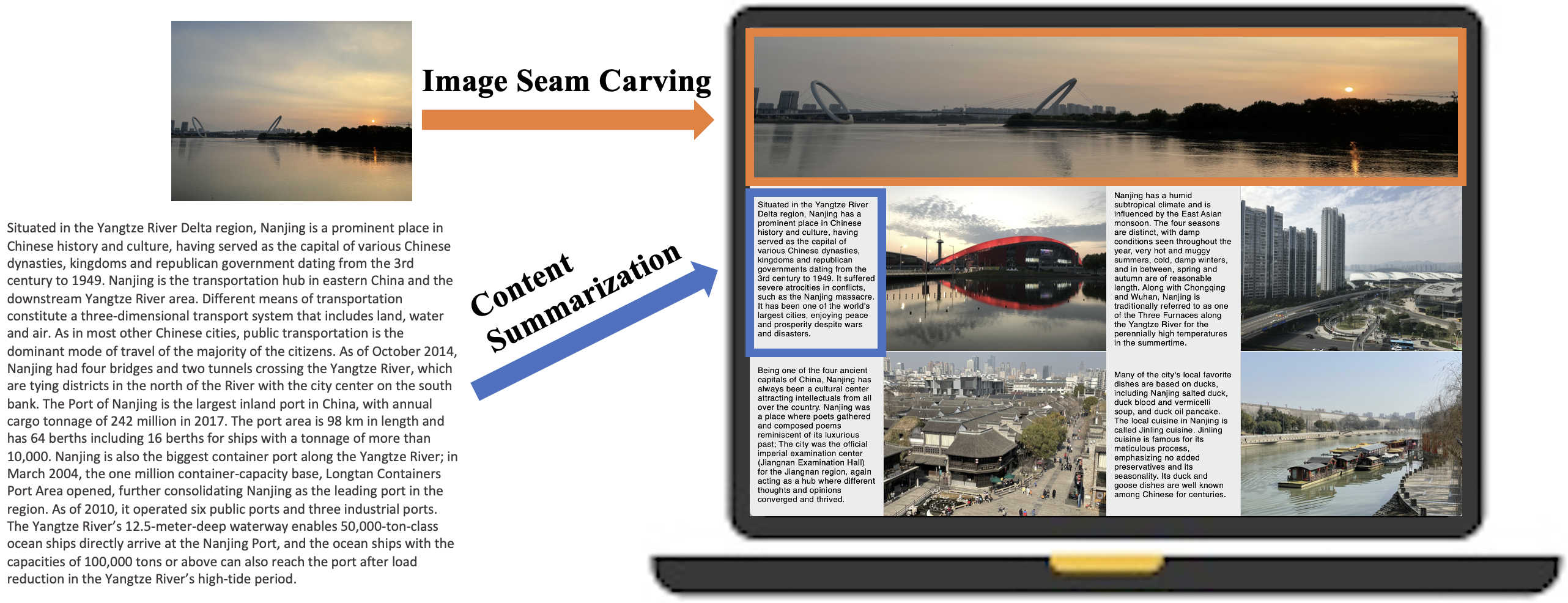}
\caption{Examples of image seam carving and content summarization.}
%\Description{Examples of image seam carving and content summarization.}
\label{fig:generation}
\end{figure}
 
Given a screen/window size, we optimize the positions and sizes of elements and the selection of alternatives through the optimization process. However, fitting all content into a layout is often challenging, especially for smaller screen sizes. One solution is to scale images and text. However, this would reduce content readability if images or fonts become too small. Alternatively, we could maximize readability by cutting content, but then we would lose (too) much information. 
 
To accommodate the vast diversity of screen sizes and document author and viewer preferences, \papername selects or dynamically generates alternative content. It applies image processing techniques such as seam carving to adapt images and BERT-based text summarization to generate alternative texts in variable sizes. It then optimizes across all the potential alternatives that conform to the given screen size and viewer preferences. For example, \papername will generate a smaller version of an image or a summarized version of a paragraph when the document is read on a small-screen device like a mobile phone. \papername also supports ``content replacement plugins'', meaning that the two implemented methods can be easily replaced with different image resizing and text summarization algorithms.

\subsection{Image Seam Carving}
 
Free-form scaling is the most common way to generate alternative images that fit a different screen size. However, standard image scaling often leads to a change in aspect ratio and thus distorts images~\cite{laine2021responsive}. An effective way to resize images based on geometric constraints considers the image content. Seam carving is a content-aware image resizing approach that supports both image reduction and expansion~\cite{aviden2007seam}, which can re-target an image to fit the expected size and aspect ratio while maintaining important content in the image and reducing unexpected distortions (Figure~\ref{fig:generation} a). \papername uses this seam carving method to generate image alternatives automatically. 

\subsection{Content Summarization}

Due to the diversity in screen sizes, a layout may not be able to include the complete text, even if vertical scrolling is enabled. Previous work~\cite{borning2000constraint} directly scaled fonts, which can significantly affect readability. Instead, it is often preferable to shorten the content so that the overall result maintains readability without losing important information. Additionally, different viewers likely have different background knowledge/interests in the same content. Some people might need more detailed information, while too detailed information might be redundant for others who are only interested in the gist of the text. It is thus helpful to have access to text alternatives with different lengths to meet these varying needs. To generate reasonable shortened versions of a piece of text, we utilize the BERT model~\cite{devlin2018bert} to automatically generate summarized versions~\cite{miller2019leveraging} (Figure~\ref{fig:generation} b). Alternatively, we could use large language models (LLMs) to generate various versions of the text, but LLMs may result in higher latency.

\subsection{Alternative Modalities}
  
Depending on screen sizes, author and viewer preferences, and requirements such as accessibility, the same information may have to be presented through alternate modalities. For example, an image may require alternative text. We currently expect document authors to provide the content for different alternative modalities. However, \papername's content replacement plugin architecture makes it easy to automate the generation of alternative modalities if suitable algorithms are available, e.g., alternative text or audio generated by machine learning models%or  For example, some machine learning models can automatically generate alternative text describing the content of an image, and text-to-speech systems could generate alternative audio
, like in Figure~\ref{fig:teaser}a. \papername then optimizes the selection of alternative content modalities. For example, to support people who prefer more visual information, \papername can use images to replace text to meet their needs. %or vice versa.
\section{Document Authoring and Viewing}
\label{sec:authoringviewing}

% \begin{figure*}
%   \includegraphics[width=\textwidth]{figures/teaser.png}
%   \caption{
%   a) \papername\ adapts a news page on a mobile phone to provide a compact overview with quick access to audio content for each article, based on (previously expressed) viewer preference (sliders below images), which prioritizes audio content over images and text here.
%   b) The same news page adapted for a tablet device with a user preference for image content.
%   c) As the viewer `pins' the COVID article and `zooms in' on the Mars article, \papername\ rearranges the layout accordingly, keeping the pinned article in place.
%   d) As the viewer `zooms in' on the blue text paragraph in the previous image with a preference for avoiding scrolling, \papername\ extends the paragraph to provide more details and crops the top image, avoiding the need for scrolling. Image credits: Production Perig/stock.adobe.com and NASA/JPL-Caltech.
%   }
% %  \Description{}
% \end{figure*}

\begin{figure*}[t!]
\centering
\includegraphics[width=\textwidth]{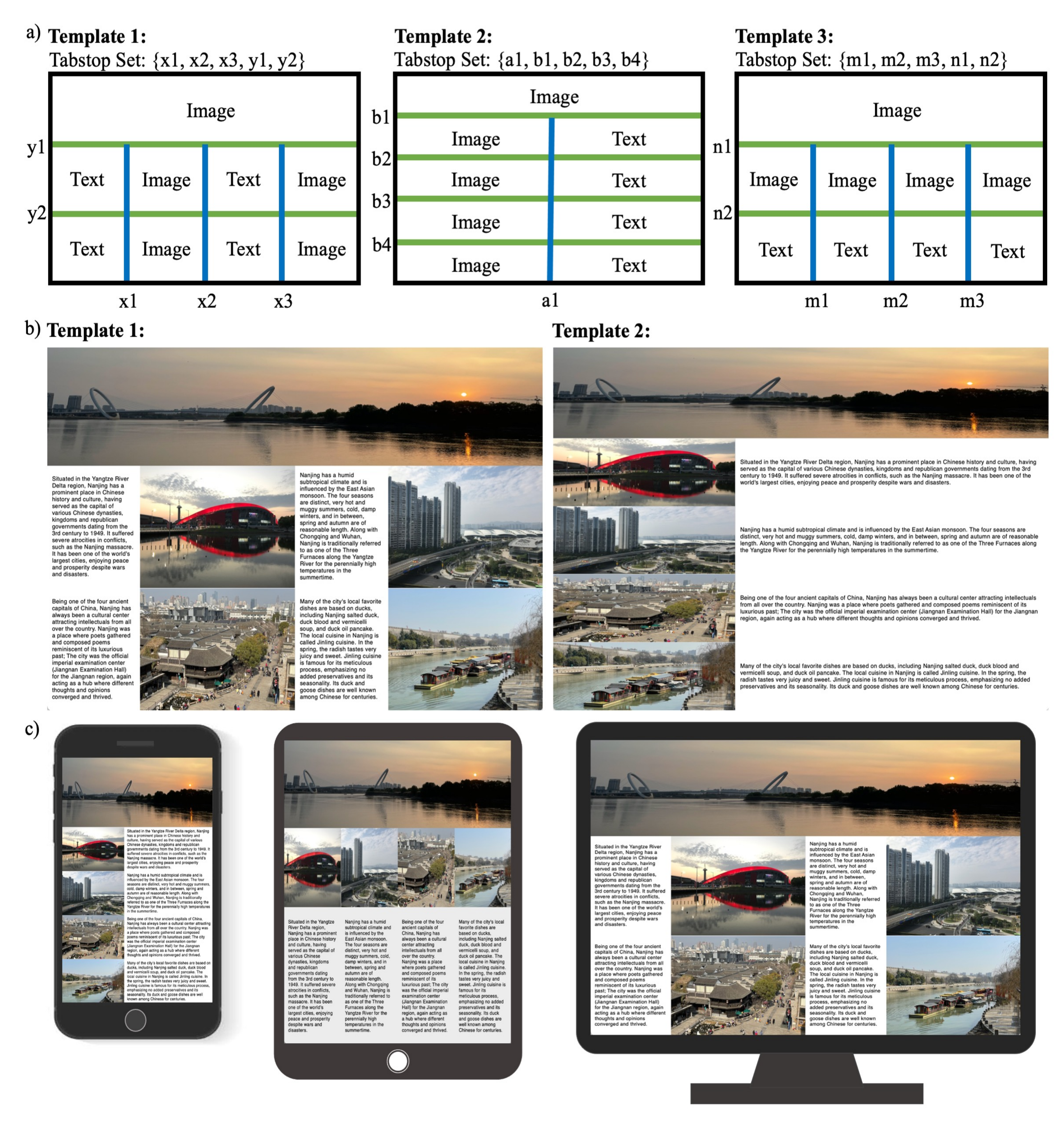}
\caption{Document optimization results: a) The author defines three templates. \papername optimizes tabstop positions based on the objective function. b) Document results when the viewer prefers the first or second template, respectively. c) Document results on different devices. The results balance both layout structure and the amount of content.
}
%\Description{This figure illustrates the document optimization process: a) Each template is defined as a set of tabstops, which are grid lines defining the boundaries of the document elements; b) shows that \papername generates new images and text by image seam carving and content summarizing; c) demonstrates that the result balances both layout structure and the amount of content.}
\label{fig:result}
\end{figure*}

% \begin{figure*}[t]
% \centering
% \includegraphics[width=\textwidth]{figures/framework.pdf}
% \caption{
% The layout template can be extracted through reverse engineering of an existing document or created directly by the author. Subsequently, the document can then be optimized based on the screen size and author preferences. Once the screen size changes and/or the viewer indicates their preferences, the document can adapt accordingly.
% }
% %\Description{}
% \label{fig:framework}
% \end{figure*}

\begin{figure*}[t]
\centering
\includegraphics[width=0.8\textwidth]{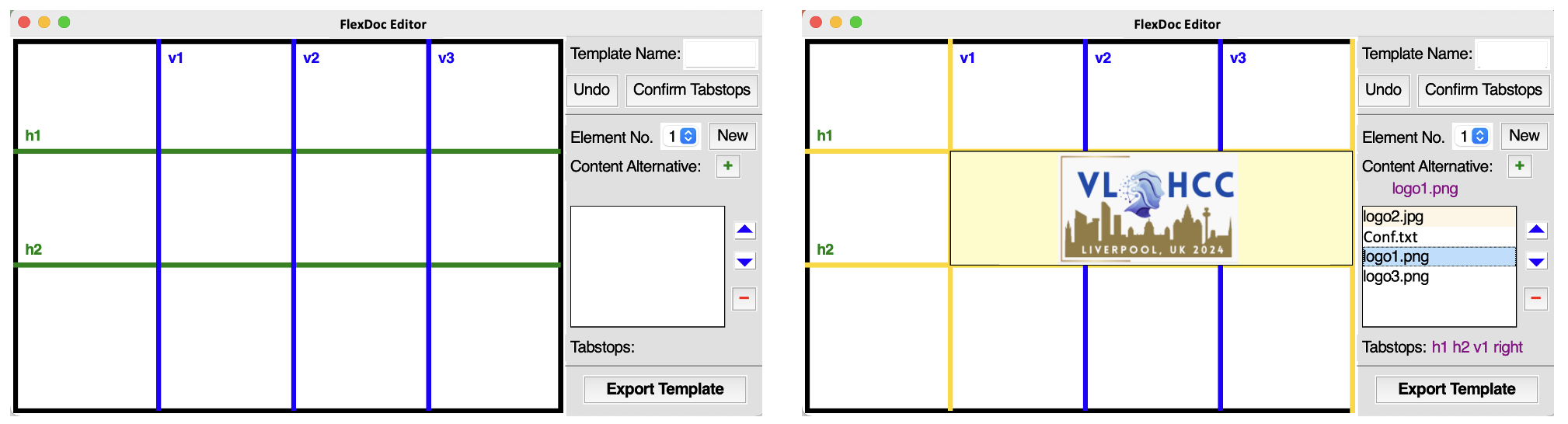}
\caption{
\papernameEditor~ is used for authoring adaptive documents. It allows authors to create templates by specifying tabstops and document elements: 
a) Horizontal (green) and vertical tabstops (blue) are created by clicking on the canvas.
b) Document elements are placed by selecting the surrounding tabstops or layout boundaries (yellow). Authors can then use the +/- buttons to add/delete alternatives for a document element and the up/down arrow buttons to specify their preference ranks.
}
%\Description{This figure shows how to author adaptive documents. a) The author uses \papernameEditor~ for template creation to define tabstops and document elements. Tabstops can be created by simply clicking on the canvas. b) Once confirmed, clicking on the canvas identifies the current document element's surrounding tabstops or layout boundaries. Authors can then upload alternatives for the target area and reorder the preference ranks.}
\label{fig:editor}
\end{figure*}

\papername~ optimizes the document based on the screen properties, author preferences, and viewer preferences. 
%As shown in \autoref{fig:framework}, the layout template can be extracted through reverse engineering of an existing document or created directly by the author.
Authors can define their content preferences using \papernameEditor, a graphical document template editor (\autoref{fig:editor}). This editor allows \emph{authors} to guide the optimization process by providing different layout templates, content alternatives, and preference rankings. 
Subsequently, the document can then be optimized based on the screen size and author preferences. 
On the other hand, \emph{viewers} can also adapt a document by selecting different layout templates or adjusting their preferences via simple operations. %sliders (e.g., for different content modalities).%, or interacting directly with the content. 
Once the screen size changes and/or the viewer indicates their preferences, the document can adapt accordingly.
These changes are saved and applied to other documents, reducing the need for repeated adjustments and ensuring a consistent viewing experience.

\subsection{Layout Templates}

% tabstops
\papername~ uses \textit{tabstop}-based layout templates to create adaptive documents due to their flexibility \cite{lutteroth2008domain}. A tabstop is a symbolic object that represents the alignments of widgets in a layout~\cite{hashimoto1992graphical, hudson1990interactive, lutteroth2006user, zeidler2012auckland, jiang2020reverseorc}. A \textit{tabstop} is essentially a variable, with an x-tabstop defining a vertical grid line through a position on the x-axis and a y-tabstop defining a corresponding horizontal grid line on the y-axis. 
%Compared to a grid-based layout, tabstops can reorder on demand if no constraints prevent this, whereas grid cells can only collapse to zero. Thus, tabstops can move through widget/content boundaries, which makes tabstops more expressive than grids \cite{lutteroth2006user}. 

% templates
A \textit{template} is defined as a set of tabstops used to align the elements that make up a document's content. Each document element is assigned to an area defined by two horizontal tabstops and two vertical tabstops. If multiple items are assigned to the same area, they flow in that area one after the other in a specified direction, similar to text. A document element may have different alternatives, with preferred alternatives having higher priorities during optimization. Templates can be predefined, reused, and shared, which reduces the burden of document authoring.

\subsection{Authoring Documents (Author Interface)}

%\subsubsection{Template Extraction through Reverse Engineering}
%To create document templates, authors can extract the layout template from existing documents through reverse engineering~\cite{jiang2020reverseorc}. We define a tabstop as a layout divider if it cleanly separates the layout into two parts without intersecting any widget within the layout. To extract the template of a document, we recursively apply the concept of layout dividers to the sublayouts contained within a layout. In the case of a horizontal layout divider, all elements are positioned either above or below it, and similarly for vertical layout dividers. The template comprises all such layout dividers.

\subsubsection{Template Creation}
Authors can also use the \papernameEditor~ to create templates. Authors can edit tabstops, element areas defined by tabstops, and preference ranks for alternative content, as illustrated in Figure~\ref{fig:editor}. Tabstops are defined by clicking on the editing canvas to set relative positions. Left clicks create horizontal tabstops (green lines), and right clicks create vertical tabstops (blue lines). Authors can then place document elements by selecting two horizontal and two vertical tabstops or layout boundaries; the selected tabstops and respective areas are highlighted in yellow. 

% alternatives
\subsubsection{Content Alternatives}
Authors can define alternatives for each document element. \papernameEditor~ supports alternatives with mixed modalities for the same element, {\em e.g.,} a document element can have both image and text alternatives. The preference ranks of alternatives can be modified via a list widget. Preferred alternatives rank higher in the list and thus get higher priority during optimization (Figure~\ref{fig:editor} b). The resulting tabstops and element alternatives with their preference ranks are then exported as a JSON file, directly used in the optimization process when viewing a document. \papername~ can also use multiple templates to optimize a single document, allowing authors to rank the templates available for a document by preference. Authors can use \papernameEditor~ to visually create templates for web documents without manually defining properties in HTML files, easing the burden of adaptive website creation. 

\subsection{Viewing Documents (Viewer Interface)} \label{sec:viewer-interaction}

%\papername~ enhances the viewing experience by personalizing documents according to the viewer's preferences. %Instead of presenting a complex user interface with document alternatives and preference rankings, 

\papername~ provides simple operations that allow viewers to express other preferences and interactively adapt documents while reading:

\paragraph{Sliders}
\papername~ allows viewers to adjust their high-level preferences using sliders %that range from 0 to 1
(Figure~\ref{fig:teaser} ab). Viewers can use these sliders to express preferences or dislikes for specific content modalities (e.g., images). \papername~ then automatically generates the corresponding objective terms, optimizing the document to align as closely as possible with the viewer's preferences. These preferences can be applied across documents to ensure a consistent reading experience. 

\paragraph{Zoom In} 
Viewers can indicate their interest in specific content by clicking on or touching it (Figure~\ref{fig:teaser} bcd). \papername~ then increases the detail of the content where viewers have shown interest. In the backend, \papername~ re-ranks alternatives for the indicated content, prioritizing those with more detail (larger images, longer text).

\paragraph{Zoom Out} 
This operation is the inverse of `Zoom In', causing \papername~ to re-rank alternatives for the content so that those with less detail are prioritized.

\paragraph{Pin} 
Viewers can pin content to its current location by double-clicking on the content element (Figure~\ref{fig:teaser} bcd). \papername~ fixes this content while optimizing other content.

\paragraph{Switch Template} 
If authors have provided multiple alternative layout templates for a document, viewers can express a preference for a specific template by selecting it from a list (\autoref{fig:result} b). If a valid solution exists, \papername~ will optimize the document using the selected layout template.

\begin{figure*}[t!]
\centering
\includegraphics[width=\textwidth]{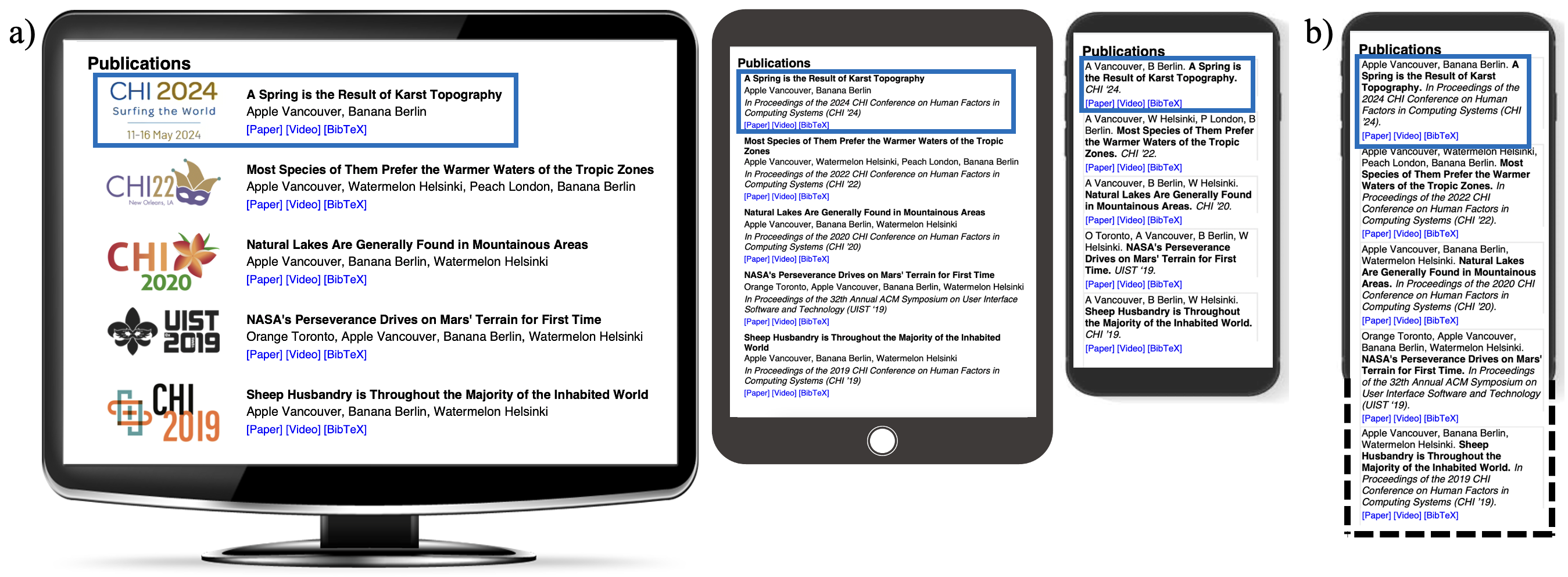}
\caption{a) \papername can generate adaptive websites with lists of publications for different devices. Corresponding alternative content is used in the different layouts (in blue boxes). For example, the conference logo (on the desktop), the full name of the conference (on a tablet), and the conference abbreviation (on a mobile) are alternatives. b) Personalized mobile version of the website if the viewer prefers to see the full names of authors and conferences.}
%\Description{a) \papername generates adaptive websites based on devices; b) Personalized mobile version of the website if the viewer prefers to show full names of authors and conferences.}
\label{fig:academic_website}
\end{figure*}

\paragraph{Switch Element} 
Viewers can switch to an alternative representation of a document element (e.g., text vs. image, longer vs. shorter text) by right-clicking on the element and selecting a preferred alternative from the available options (\autoref{fig:academic_website} b). Then, \papername~ optimizes the document based on the selected option, provided a valid solution exists.

\section{Applications}
\label{sec:application}

We demonstrate \papername in multiple real-world application scenarios. In addition to the new website shown in the main paper, we show three additional examples.

% \begin{figure*}[t!]
% \centering
% \includegraphics[width=\textwidth]{figures/new_result.png}
% \caption{Document optimization results: a) The author defines three templates. \papername optimizes tabstop positions based on the objective function. b) Document results when the viewer prefers the first or second template, respectively. c) Document results on different devices. The results balance both layout structure and the amount of content.
% }
% %\Description{This figure illustrates the document optimization process: a) Each template is defined as a set of tabstops, which are grid lines defining the boundaries of the document elements; b) shows that \papername generates new images and text by image seam carving and content summarizing; c) demonstrates that the result balances both layout structure and the amount of content.}
% \label{fig:result}
% \end{figure*}

\begin{figure*}[t!]
\centering
\includegraphics[width=\textwidth]{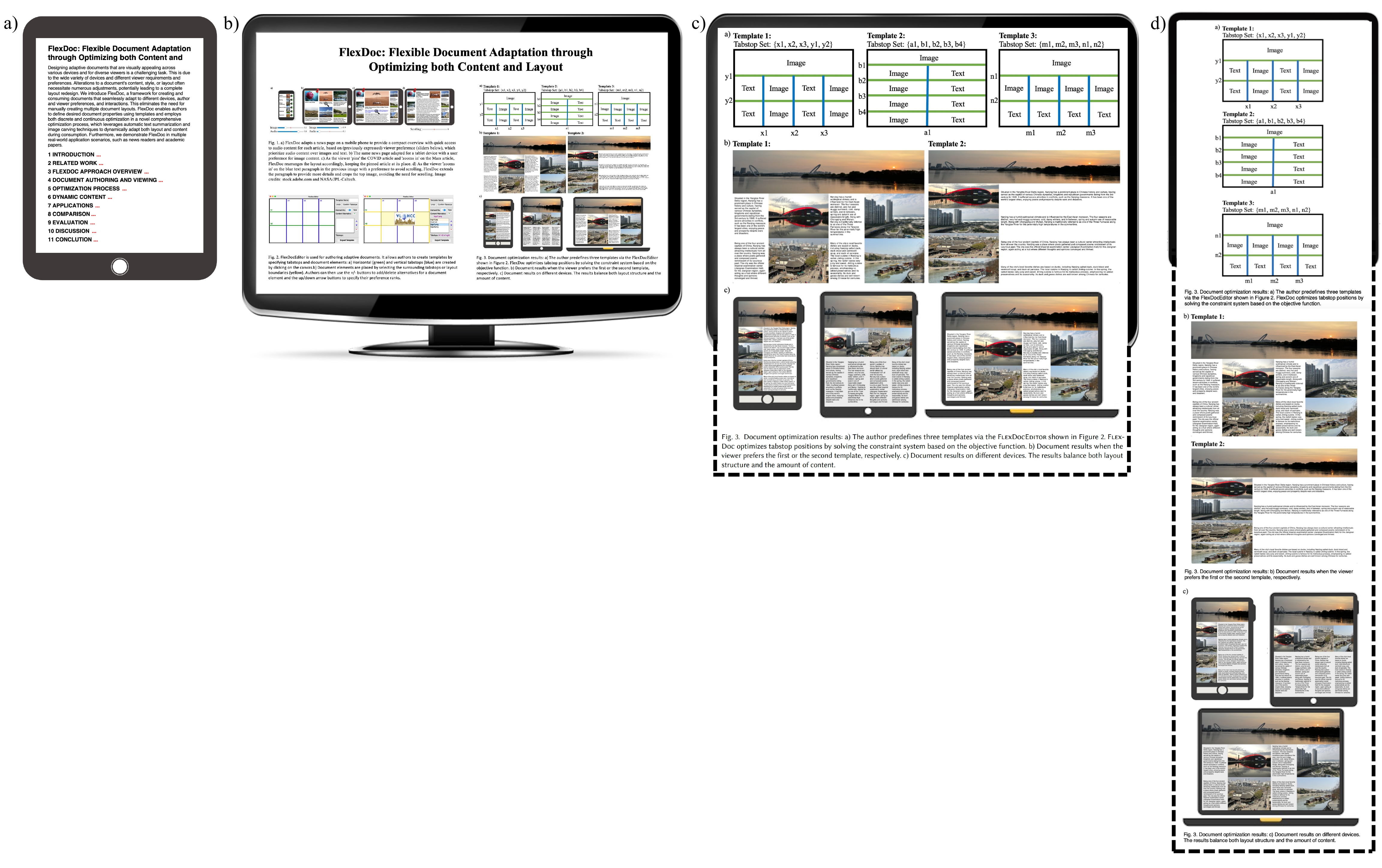}
\caption{An application of \papername for adapting the layout of scientific papers based on the screen size and author and viewer preferences. The figure illustrates the results of applying \papername to our \papername paper in various scenarios: a) The viewer prefers to see only sections, for instance, to gain an overview. b) The viewer prefers a figure-only two-column format of the paper that fits the screen, for instance, to gain a visual overview. c) A wide figure is shown on a desktop screen, including the option to scroll down, for instance, to inspect it in detail. d) The same figure is shown on a small screen with automatic reflowing, including the option to scroll down.}
%\Description{This figure demonstrates that \papername generates adaptive scientific papers based on device constraints and author and viewer preferences. b) a figure-only two-column format of the paper that fits the screen size; c) a wide figure is shown on a desktop screen (including the option to scroll down); f) the same figure shown on a small screen by reflowing automatically (including the option to scroll down).}
\label{fig:paper}
\end{figure*}

\subsection{Scientific Paper}

Reading scientific papers on devices with small screens, especially when they contain wide figures, can be challenging. Additionally, when reading papers, people often have to scan the entire document to find relevant sections they want to read first. Some individuals prefer to start by reading the abstract and the body text, while others might prefer to begin by scanning the figures and their captions to get a general idea of the paper. Furthermore, the ability to automatically convert a paper into different formats can significantly save time for scientific authors and help viewers better understand the content.

\papername addresses these use cases by adapting a paper according to device properties and viewer preferences (see Figure~\ref{fig:paper}). For instance, if the viewer prefers to see only sections to gain an overview, \papername can generate a version with section titles only (Figure~\ref{fig:paper}a). If the viewer prefers a figure-only two-column format of the paper that fits the screen to gain a visual overview, \papername can generate a version with figures only (Figure~\ref{fig:paper}b). On a desktop screen, a wide figure can be displayed. The same figure is shown on a small screen with automatic reflowing, allowing for scrolling down.

\subsection{Academic Website}

\papername can generate adaptive websites with lists of publications for different devices by selecting the most suitable content alternatives based on device properties and viewer preferences. Corresponding alternative content is used in the different layouts, highlighted in blue boxes. For example, the conference logo is displayed on desktops, the full name of the conference is shown on tablets, and the conference abbreviation is presented on mobile devices (see Figure~\ref{fig:academic_website}a). If the viewer prefers to see the full names of authors and conferences, \papername can generate a personalized mobile version of the website by allowing the viewer to switch elements. This can be done by right-clicking on the element and selecting a preferred alternative from the available options.

\subsection{News Website with Advertisements}

\begin{figure*}[t!]
\centering
\includegraphics[width=\linewidth]{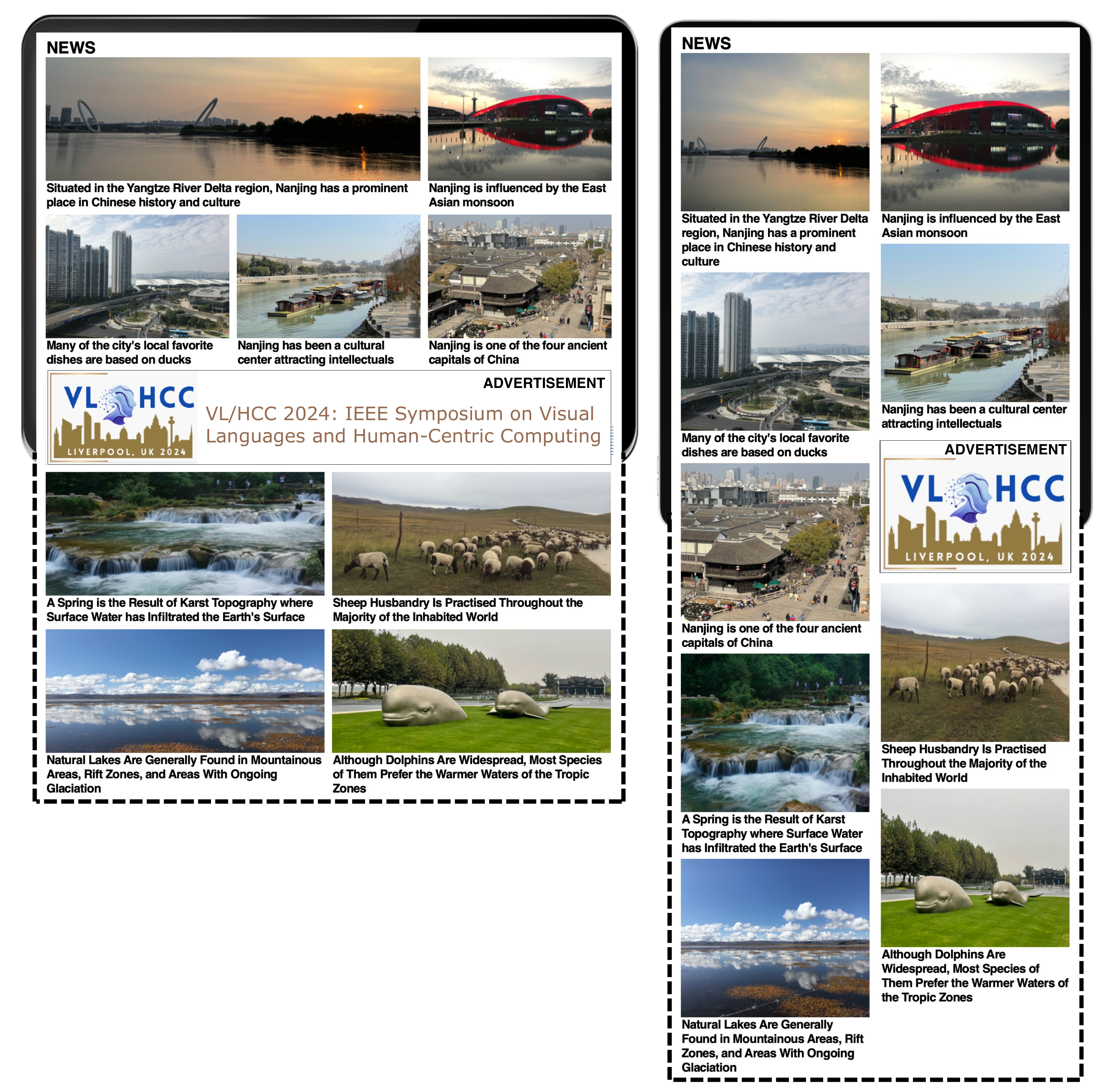}
\caption{A news website example on different devices with a ``VL/HCC2024'' advertisement, including the option to scroll down. On the desktop, all the rows are aligned while on a tablet, columns are aligned. Different alternatives to the ``VL/HCC2024'' advertisement are selected based on the device properties and the layout structure to avoid the advertisement occupying too much space while still drawing attention.}
%\Description{A news website example on different devices with a ``DIS2024'' advertisement.}
\label{fig:website_ads}
\end{figure*}

We present an additional news website example containing a ``VL/HCC2024'' advertisement to demonstrate how \papername can adapt documents to different devices (Figure~\ref{fig:website_ads}). 
 On the desktop, all the rows are aligned, while on a tablet, columns are aligned. Different alternatives to the ``VL/HCC2024'' advertisement are selected based on the device properties and the layout structure to avoid the advertisement occupying too much space while still drawing attention.

%Selecting the right length of a text then involves a trade-off between content loss and readability, which can additionally also depend on viewer preferences.

%Similarly, if the viewer indicates interest by clicking on or touching content (`Zoom In'), alternatives with more detail than the current alternative (e.g., larger images or longer text) are given a large weight, while other alternatives for the same content element get a weight of $0$.

\begin{table*}[t]
\def\w{0.18\linewidth}
\def\ww{0.4\linewidth}
 \centering
\begin{tabular}{ | c | c | c | c |}
 \hline
Input Document  &  Adobe Liquid Mode & Webflow & Ours \\
\hline
 & & & \\
\includegraphics[width=\ww, valign=t]{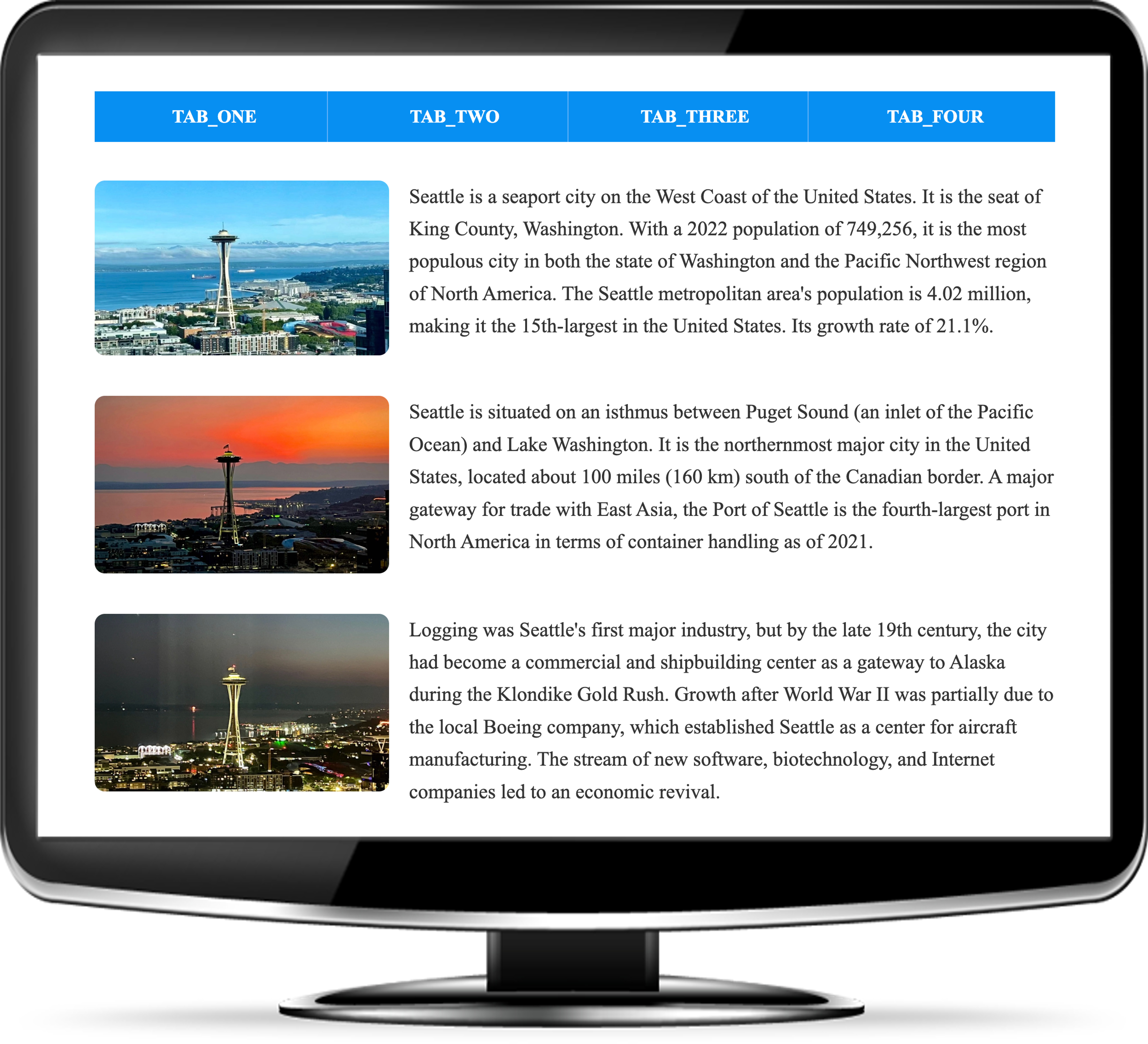} &
\includegraphics[width=\w, valign=t]{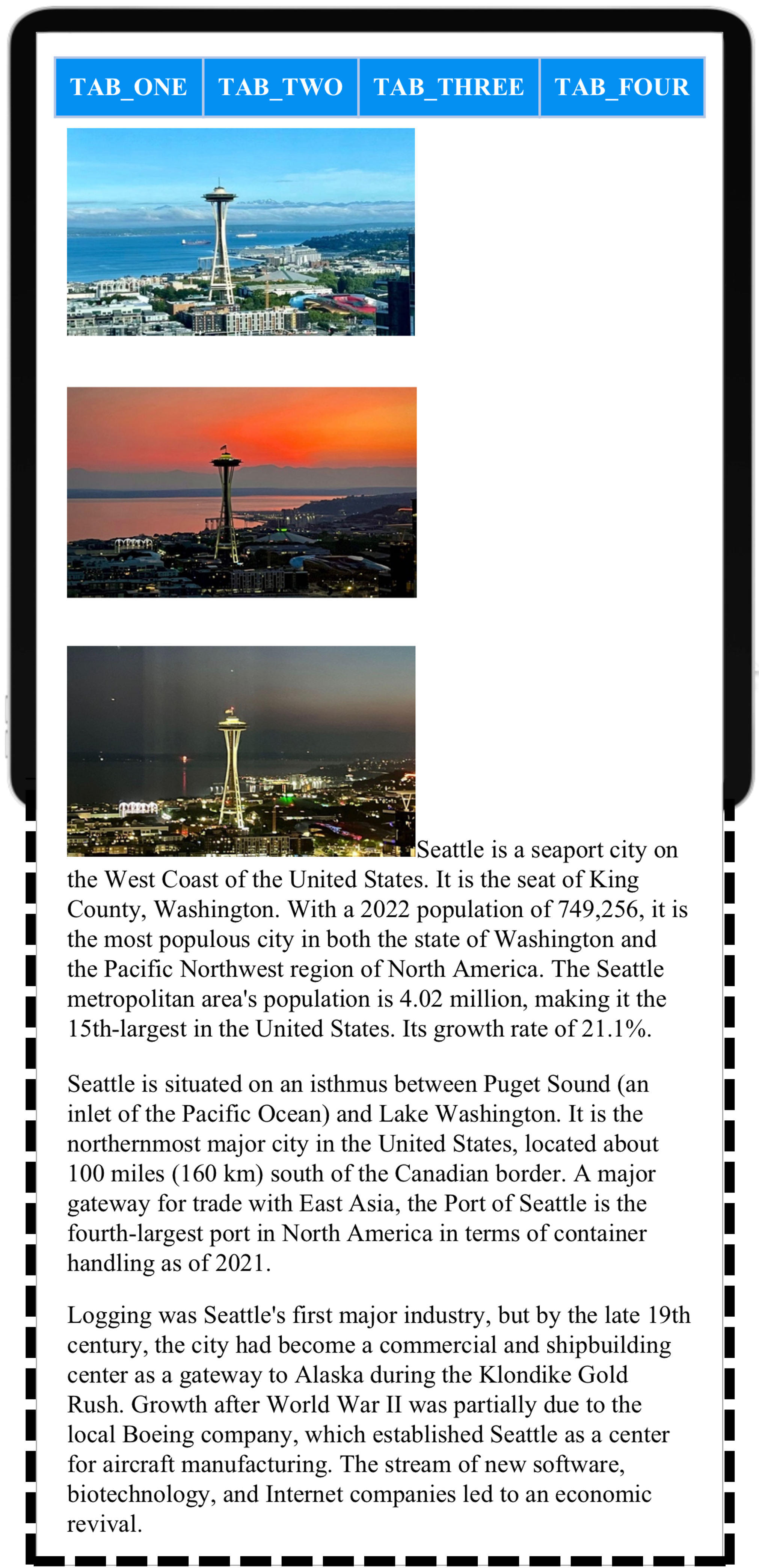} &
\includegraphics[width=\w, valign=t]{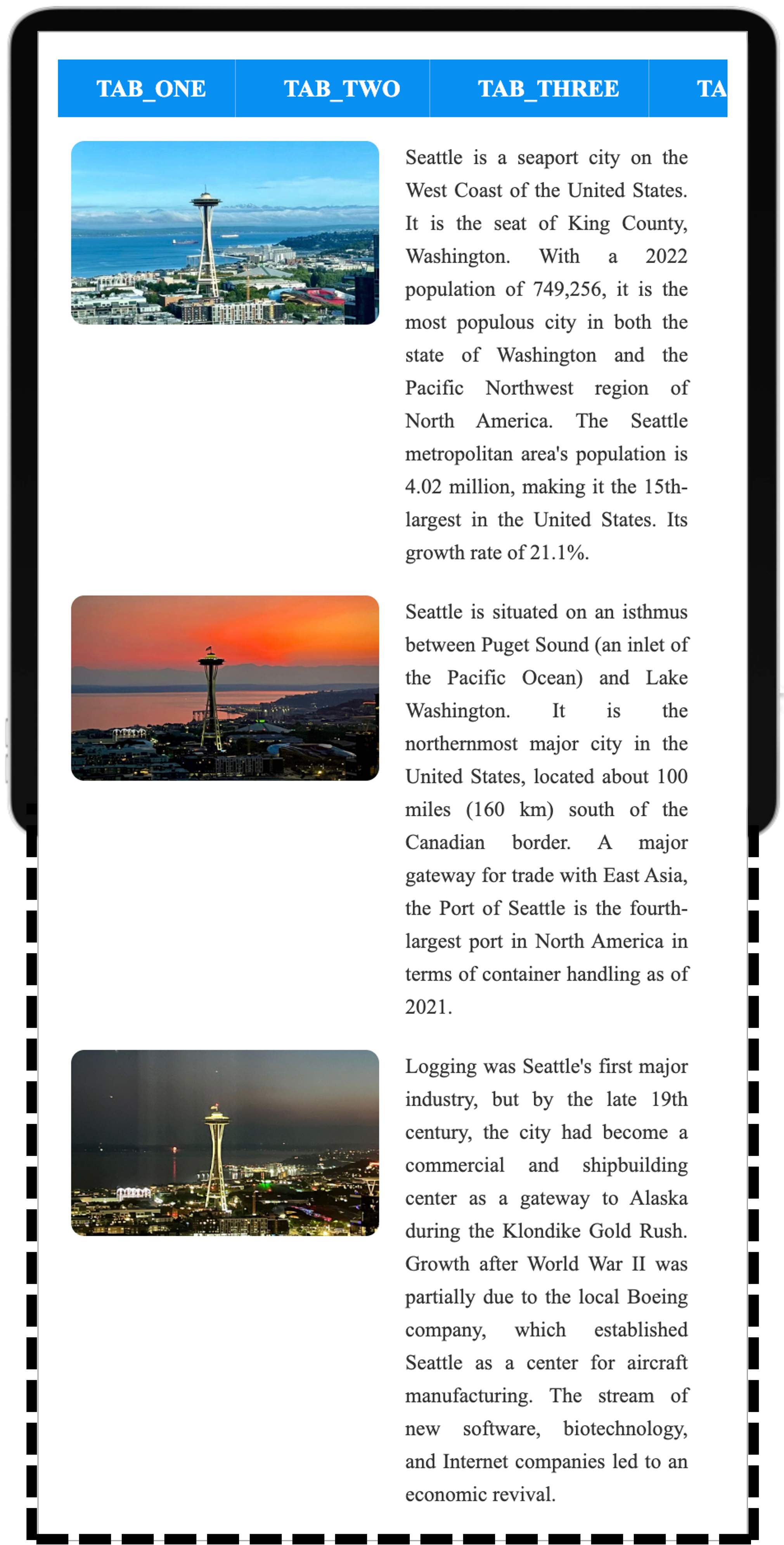} &
\includegraphics[width=\w, valign=t]{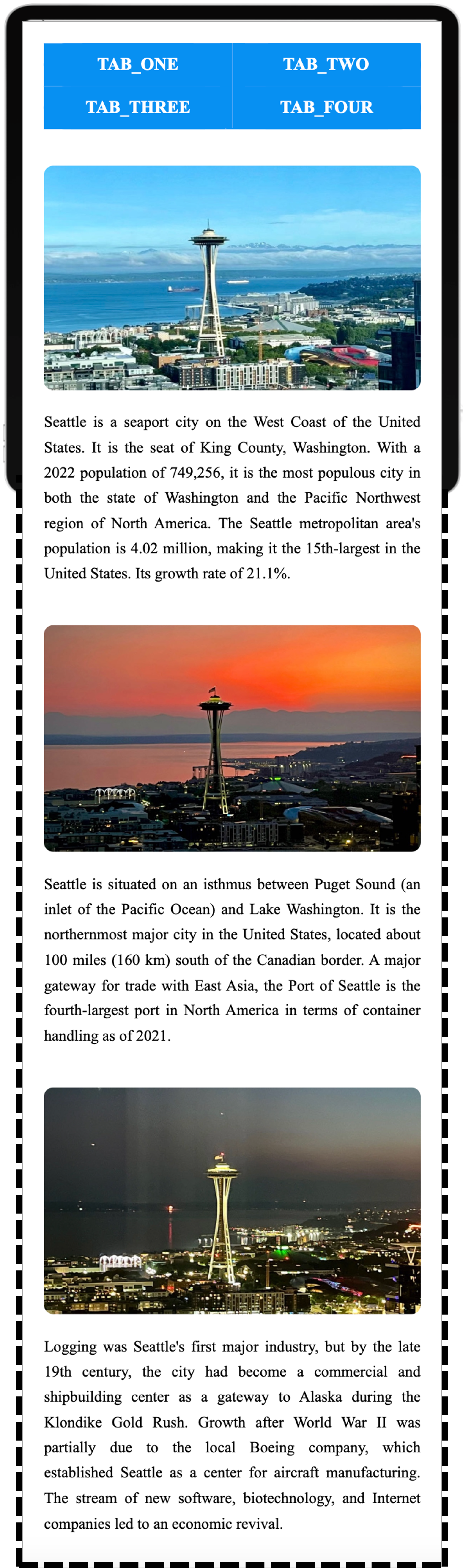}\\
 & & & \\
 \hline
\end{tabular}
\caption{Comparison between our algorithm and other existing tools, in a scenario where when we adapt a document on the desktop to a smaller screen.}
\label{fig:comparison}
\end{table*}

\section{Comparison}

We illustrate the comparison between our algorithm and other existing tools in adapting a desktop document to a smaller screen in Table~\ref{fig:comparison}.
Our \papername algorithm is capable of generating adaptive content, enabling image flow, and producing adaptive documents with multiple columns, all without the need for breakpoints (screen sizes at which the document layout changes) or coding. In contrast, Adobe Acrobat Liquid Mode is restricted to a single-column format for mobile document viewing, lacks automatic adaptation to a multi-column format, and does not support image flows.
Responsive Web Design (RWD) utilizes HTML and CSS to automatically resize, hide, shrink, or enlarge a website to fit different devices. However, it requires coding and places the onus on authors to manually code and set breakpoints for different layout sizes. While Webflow eliminates the need for coding, it still requires authors to establish breakpoints for varying screen sizes and does not support image flows without designing different layouts for different devices.
Furthermore, all these approaches, except our \papername, fail to generate adaptive content. This means that once the design is finalized, the content becomes fixed, thereby limiting viewers from modifying the content according to their preferences. \papername, on the other hand, can adapt to different viewer preferences, a feature currently unavailable in other tools.

In addition, we compare \papername to previous adaptive document layout methods.
In contrast to Zooming User Interfaces (ZUIs), e.g., \cite{bederson1996pad++}, which make some components invisible while zooming, \papername enables adaptive content generation and optimizes layouts to emphasize the object of interest while also keeping other components visible/readable and the layout reasonably close to the original one. 
Compared to content management systems (CMSs), \papername provides more expressiveness for authors and a more personalized experience for viewers. \papername could be integrated into a CMS to address the limitations of such systems.
Responsive web design (RWD) uses HTML and CSS to automatically resize, hide, shrink, or enlarge a website to fit different devices. Compared to RWD, \papername can adapt to different viewer preferences, which is not currently possible for RWD.

\section{Use Cases}

%Our proposed \papername approach dynamically optimizes both the structure and content of a document to adapt it to various device properties and the preferences of authors and viewers. \papername can be applied to a wide range of document-centric applications, enhancing both the reading and authoring experience across various document types. Designers can use \papernameEditor~ to create flexible document layout templates that ensure readability across different use cases. \papername can then generate suitable versions of both images and text content to fit these different layout templates. 

%\subsection{Customizability of Document Layout and Content}

\papername's ability to adapt to different devices and content enables adaptive documents in various use cases:

\subsubsection{Adaption to Different Devices}

Most current documents are static. Once a document has been created, viewers have very limited options to change how they view the document. For example, many documents cannot be easily shown on small screens such as mobile phones. Generating and adapting documents to different devices with various sizes, aspect ratios, and orientations is a challenging problem. In contrast, documents made with \papername can adapt more easily to different devices.

\subsubsection{Accessibility}
\papername benefits people with different document viewing requirements. It generates documents with more images when the user prefers visual content over text. \papername can also generate documents with different font sizes or even audio options if the user has vision impairments. Furthermore, alternative versions of a text with different languages can be used to support internationalization. 

\subsubsection{Task-Specific Customization}
\papername enables task-specific customization depending on viewers' needs. For example, viewers often prefer first an overview of the latest news and then consume some of the news items in more detail. \papername supports the display of content at different levels of detail, showing more detailed versions whenever the viewer indicates their interests via clicking or touching the screen.

\section{Limitation and Future Work}

\papername employs automated content generation methods to produce suitable images and texts to fit the document better. However, these methods need to be evaluated for quality and efficiency. Sometimes, a text summarization model may generate low-quality text~\cite{kryscinski2019neural}, or an intelligent seam carving approach may distort images or create visual artifacts. Nevertheless, any content generation method can be used within \papername. We provide an approach for ``content replacement plugins'' within \papername, and the two methods implemented in our prototype can be easily replaced with any desired image resizing and text summarization algorithms. Poor content generation can be identified by the author and manually addressed by replacing content accordingly. Alternatively, automated feedback, such as an ML critic model, could be used to evaluate the quality of content alternatives.

While \papername reduces the authoring effort for templates, we are currently limited to ones that can be defined via tabstops. Future work could enable the adaptation to infographics or images with irregular boundaries to different screens and user preferences. Also, our current work does not consider semantic categories or the hierarchical structure of documents. Future work could explore more extensive optimization based on the semantics of document elements and the relationships among those elements, particularly hierarchical structures.

 \section{Conclusion} 

We introduced \papername, an innovative adaptive document approach that facilitates dynamic optimization of both content and layout structure. This optimization is not solely dependent on screen or window sizes, but also takes the preferences of both authors and viewers into account. \papername optimizes the layout structure using an objective function, condenses content to accommodate the layout, and further allows the outcome to adapt based on user interactions. We anticipate that our method could have broad applications across diverse screen sizes and document types. 
The versatility of \papername paves the way for potential future document-centric applications. 
Additionally, future work can move beyond the current model where authors predefine extended content or extract background information from previous articles. Instead, we could develop more sophisticated approaches to dynamically extend documents, and generate more relevant, high-quality details.
For example, in future work content generation approaches such as Large Language Models (LLMs) could be used as part of the creation of intelligent, flexible documents to provide suitable content for layouts that are in real time optimized to fit the available space and context.

\bibliographystyle{ieeetr}
\bibliography{Reference}

\end{document}